\PassOptionsToPackage{table}{xcolor}
\documentclass[acmsmall]{acmart}

\setcopyright{acmlicensed}
\copyrightyear{2018}
\acmYear{2018}
\acmDOI{XXXXXXX.XXXXXXX}

\usepackage{pdflscape}
\usepackage{microtype}
\microtypesetup{expansion=false}

\usepackage{natbib}
\usepackage{tikz}
\usepackage{pgfplots}
\pgfplotsset{compat=1.18}
\usepackage{enumitem}
\usepackage{multirow}
\usepackage[edges]{forest}
\usetikzlibrary{arrows.meta,positioning,shadows.blur,fit}

\forestset{
  mybox/.style={
    draw=gray,  
    rounded corners=3pt,
    fill=gray!5,
    inner sep=3pt,
    outer sep=0pt,
    text width=52mm,      
    align=center,
    font=\footnotesize,   
    drop shadow={blur radius=1pt, opacity=.25}
  },
  myedge/.style={line width=0.7pt, draw=gray},
}

\acmJournal{JACM}
\acmVolume{37}
\acmNumber{4}
\acmArticle{111}
\acmMonth{8}

\begin{document}

\title{Toward Intelligent Prefetching: A Survey on Complex Memory Access Prediction Techniques}

\author{Sheel Sindhu Manohar}
\authornote{Both authors contributed equally to this research.}
\email{sheel.manohar@snu.edu.in}
\orcid{0000-0001-7490-6209}
\affiliation{%
  \institution{Shiv Nadar IoE}
  \city{Greater Noida}
  \state{UP}
  \country{India}
}

\renewcommand{\shortauthors}{Manohar et al.}

 
\begin{abstract}
Data prefetching is a critical technique for bridging the processor-memory performance gap by predicting future memory accesses and retrieving data into on-chip caches before demand. While traditional prefetchers based on next-line, stride, and correlation heuristics perform well for regular access patterns, they are fundamentally inadequate for the irregular, data-dependent patterns prevalent in modern workloads such as graph analytics, sparse matrix computations, and pointer-intensive applications. This survey presents a systematic review of papers using a PRISMA-guided selection methodology. We propose a structured taxonomy that organizes prefetching techniques across three dimensions: locality type (spatial vs.\ temporal), implementation layer (hardware, software, and hybrid), and--for the increasingly important class of ML-based prefetchers---learning paradigm (supervised, reinforcement, and unsupervised) paired with training mode (online vs.\ offline). Through a multi-dimensional comparative analysis of ML-based prefetchers evaluated across storage overhead, accuracy, inference latency, hardware feasibility, and generalization ability, we identify three key findings: (i)~an accuracy--overhead Pareto frontier defined by model class, (ii)~a natural architectural mapping between model complexity and cache hierarchy level, and (iii)~a fundamental tension between runtime adaptability and model capacity that motivates hierarchical ensemble architectures. We further enumerate concrete research gaps organized across seven categories, including the absence of standardized ML-prefetcher benchmarks, and LLM inference workloads, uncharacterized energy--accuracy trade-offs, and the lack of theoretical frameworks for analyzing prefetchability. This survey provides researchers and system designers with both a comprehensive reference for current techniques and a structured roadmap for future investigation.
\end{abstract}

\ccsdesc[500]{Do Not Use This Code~Generate the Correct Terms for Your Paper}
\ccsdesc[300]{Do Not Use This Code~Generate the Correct Terms for Your Paper}
\ccsdesc{Do Not Use This Code~Generate the Correct Terms for Your Paper}
\ccsdesc[100]{Do Not Use This Code~Generate the Correct Terms for Your Paper}

\keywords{data prefetching, cache prefetching, memory access prediction, machine learning, reinforcement learning, neural prefetcher, spatial prefetching, temporal prefetching, hybrid prefetching, survey}

\received{20 February 2007}
\received[revised]{12 March 2009}
\received[accepted]{5 June 2009}

\maketitle

\section{Introduction}
Data prefetching is a critical technique for enhancing computational performance by reducing memory latency, particularly in high-performance computing systems. Prefetching predicts the data the processor will require and retrieves it from main memory into on-chip caches in advance~\cite{FDP,OldPrefetcher2}. The emergence of teraflop-scale computing on many-core systems has triggered a significant increase in complexity~\cite{FDP,TeraFlop1,TeraFlop2,TeraFlop3,TeraFlop4,TeraFlop5,TeraFlop6,TeraFlop7}. The workloads generated by various applications require a more effective caching mechanism on the chip. However, modern CPUs have a caching mechanism that exploits locality of references; we still need to deploy multiple prefetchers across various cache levels to enhance overall performance~\cite{b1}. The prefetcher accurately anticipates future data, resulting in a significant reduction of long main memory access latency.~\cite{memorywall} Thus, we get an improved performance in terms of throughput~\cite{ref1,ref2}. 
Nowadays, complex applications possess irregular data structures that exhibit reduced spatial locality and require efficient prediction mechanisms. Recently, researchers focused on the underlying issues of unpredictable access patterns. Current research emphasizes intelligent prediction mechanisms, particularly while maintaining minimal design complexity~\cite{challenge7,challange2}. Traditional prefetching techniques, such as next-line, stride, and correlation-based schemes, are effective for workloads with regular or repetitive patterns~\cite{OldPrefetcher1,OldPrefetcher2}. However, these methods are inadequate for irregular patterns and are application-dependent, whereas tracing changing memory behavior is now common. For example, \textbf{pointer chasing, graph traversals, or sparse matrix computations} are complex techniques that do not rely on static rules. Conventional prefetchers cannot capture higher-order relationships across spatial and temporal dimensions. Therefore, they often experience \textbf{low coverage of data, cache pollution, and bandwidth waste due to irregularity}~\cite{STMS,pythia,challange1}. 

Machine learning(ML) represents a logical progression, as it enables prefetchers to learn complex correlations directly from access streams. This approach does not require explicit programmer or compiler intervention~\cite{ref21}. ML-based approaches include~\textbf{neural networks}~\cite{ref25},\textbf{perceptrons}~\cite{ref23}, \textbf{reinforcement learning agents}~\cite{pythia}, and \textbf{decision trees}~\cite{ref11}. Static heuristics are unable to detect non-linear, and context-dependent patterns. These techniques can improve prediction algorithms with consistent feedback, generalize across a variety of inputs, and dynamically adjust to workload behavior~\cite{ref2,ref4}. \textbf{Semantic-based }prefetching approaches, which search for data context within accessible blocks and greatly increase prediction accuracy, are another advancement. Techniques like SeLeP~\cite{SeLaP} have shown enhancements in hit ratios and decreases in I/O times through the application of deep learning models. The methodologies have evolved from predictive to intelligent prefetching, whereas the system learns and adapts to intricate patterns instead of depending exclusively on historical data. The method improves prefetcher capacity to manage intricate access patterns prevalent in contemporary applications, thus overcoming the shortcomings of earlier prefetchers~\cite{ref4,ref5,ref6,ref8}. 

Prospects of using \textbf{machine learning (ML)} to estimate complex, non-uniform access patterns are one of the major and fast-expanding areas of intelligent prefetching~\cite{Delta-LSTM, ref22}. The prediction capability of ML-based prefetchers is also workload-dependent, and the insights learned can be applied to varying inputs from the program~\cite{ML1,ML2}. The advantage is the ability to make predictions with better accuracy and coverage without the involvement of a human or fixed annotations. Consequently, prefetching becomes more flexible and able to handle any type of workload and is more dependable over practical settings~\cite{ref11}. Although machine learning-based prefetchers have substantially improved prediction accuracy, their practical effectiveness remains constrained by highly irregular memory access patterns. Addressing this limitation demands a deeper understanding of the complexities underlying the prediction of such non-uniform access sequences~\cite{Delta-LSTM,Temporal4}.

\textbf{\textit{ Scope of the article}}: This article reviews contemporary research on CPU cache prefetching, focusing on advancements in handling complex data access patterns.  We specifically exclude GPU prefetching (though applicable) and other cache optimization techniques.  Prioritizing conceptual clarity over quantitative results (due to varied evaluation platforms), we categorize prefetching strategies by core characteristics, providing a concise yet comprehensive overview of recent progress for researchers, system designers, and developers. In the following article, we focus mostly on rationale thinking-based prefetching techniques, examining their design principles, strengths, and limitations in addressing non-uniform and irregular access patterns.

\begin{table*}[ht]
\centering
\scriptsize
\caption{Mapping of Prefetching Challenges to Representative Techniques}
\label{tab:challenge_technique_mapping}
\renewcommand{\arraystretch}{1.3}
\begin{tabular}{|p{3.5cm}|p{2.5cm}|p{7cm}|}
\hline
\textbf{Challenge} & \textbf{References} & \textbf{Brief Description} \\
\hline

Irregular Access Pattern Recognition
&
\cite{challange1}, \cite{challange2}, \cite{voyager}, \cite{challenge3}, \cite{challenge4}, \cite{challenge5}, \cite{challenge6}
&
Prefetchers leveraging multiple program features and cache hierarchy levels to handle branch-dependent, pointer-chasing, and irregular address streams. \\
\hline

Correlation Between References
&
\cite{STMS}, \cite{Sandbox}, \cite{Temporal6}, \cite{Temporal4}, \cite{ref9}, \cite{challenge7}
&
Multi-stream monitoring within a physical page; history buffers (RPT) and address-pair correlation (ISB) to predict correlated, non-sequential accesses. \\
\hline

Resource Overhead
&
\cite{FDP}, \cite{PPF}, \cite{b7}, \cite{hermes}, \cite{powerConsumption}, \cite{CLIP}, \cite{CDF}, \cite{BIP}, \cite{Triage}, \cite{pythia}
&
Metadata tables, on-chip buffers, ML-based models, and multilevel indexing incur area, static/dynamic power, and critical-path overhead; mitigated via compressed metadata and off-chip storage. \\
\hline

Performance Trade-offs (Accuracy vs.\ Coverage vs.\ Bandwidth)
&
\cite{FDP}, \cite{b7}, \cite{hermes}
&
Optimizing prefetch accuracy at the cost of coverage degrades performance; prefetchers must identify complex address-stream patterns within execution-time constraints. \\
\hline

\end{tabular}
\end{table*}
\subsection{Challenges in prefetching access pattern}

In order to predict, an irregular access patterns require more complex prediction algorithms. We have to face several challenges. The effective prefetching explore prefetchers that leverage multiple program features within the application and consider the cache hierarchy level.~\cite{challange1,challange2,voyager}

\textbf{Emergence of Multi-core Architectures:} In order to get the collective performance from the multiple cores, we have introduced the multiprocessor architecture.~\cite{Ghose2025} This increase in the core has also escalated the demand from the processor. To serve multiple requests from the higher-level cache, there is an increase in traffic on the critical path from main memory.~\cite{memorywall}If we receive a request with sequential memory accesses, then the prediction would always be deterministic. As system architects, we have proposed various prefetch strategies to mitigate the penalty associated with memory access.~\cite{ref1} As emerging applications become increasingly complex, they exhibit dynamic memory access patterns.~\cite{challenge3,challenge4} Thus, it would have become difficult to predict access via traditional hardware prefetching because the application primarily depends on branches, which chase data structures. So, prefetches have to deal with the irregular and complex pattern recognition.~\cite{challenge5,challenge6}

\textbf{Correlation between references:} The cache memory gets requests from the higher-level caches. In order to predict closer address profiling based on the access pattern, the prefetcher must have the capability to monitor multiple streams from memory within a single physical page.~\cite{STMS,Sandbox} To maintain these streams as a reference, one needs to store their history in some memory component.~\cite{RPT} Additionally, it must ensure that the new access address is within the previous addresses within a defined window(effectively comparing pairs of addresses)~\cite{MISB,Temporal4}, Storing correct addresses as a reference and using the history buffer judiciously is a next-level challenge.~\cite{challenge7}  

\textbf{Resource Overhead:} Traditionally, past references considered for efficient prediction often require metadata storage of prediction tables and dedicated buffer storage.~\cite{FDP} Implementation of such hardware logic incurs significant area and power overhead. Eventually, prefetcher designs emerging over the last few years, especially those using advanced prediction algorithms, machine learning techniques, or multilevel indexing schemes, tend to include extra overhead in storage complexity.~\cite{PPF,b7,hermes} For example, history-based prefetchers are susceptible to large on-chip metadata buffers or tables, which increases both static and dynamic power consumption.~\cite{powerConsumption} In some cases, prediction models incur computational overhead, which may extend cache critical paths and adversely affect clock frequency.~\cite{CLIP,CDF,BIP} Recent work mitigates these overhead issues by optimizing metadata using compressed representation schemes and intelligent off-chip metadata storage methods~\cite{Triage,pythia}, but balancing prediction accuracy with resource limitations remains a key research consideration.

\textbf{Performance Tradeoffs:} If we achieve accuracy at the expense of data coverage, we are left with poor performance. Thus, negotiating accuracy with the coverage and memory bandwidth affects the performance.~\cite{FDP} To improve both coverage and accuracy, prefetchers must be able to identify complex patterns in the address stream within the stipulated execution time.~\cite{b7,hermes} In summary, while modern prefetching techniques have made significant progress in addressing irregular access patterns, they continue to face inherent limitations arising from complex correlations, high resource overhead, and unavoidable performance trade-offs. These unresolved challenges underscore the importance of a systematic review of current approaches and motivate the need to clearly delineate the scope of this article.

\begin{table*}[h]
\centering
\scriptsize
\caption{Common Metrics and Features for Prefetcher Evaluation}
\begin{tabular}{|p{2cm}|p{5.8cm}|p{5cm}|}
\hline
\textbf{Metric} & \textbf{Definition} & \textbf{Insight} \\ \hline

\textbf{Coverage} & Proportion of cache misses reduced by prefetched cachelines.~\cite{challange1} & Indicates how effectively a prefetcher mitigates compulsory cache misses. Higher coverage means more useful prefetching. \\ \hline

\textbf{Timeliness} & Measures whether prefetched data arrives before the demand access. & Ensures data is available when needed, avoiding stall cycles.~\cite{timeliness} \\ \hline

\textbf{Lookahead} & Extent to which a prefetch is sent in advance of demand access.~\cite{MLOP} & Balances being early enough (data ready) but not so early that data gets evicted before use.~\cite{ref36,FDP,ref37} \\ \hline

\textbf{Overprediction} & Occur when prefetched data is already present in the cache. & Identifies wasted bandwidth and energy overhead.~\cite{ref39} \\ \hline

\textbf{Aggressiveness} & Degree to which the prefetcher stays ahead of the demand stream and the number of requests generated. & Too aggressive $\rightarrow$ cache pollution \& bandwidth waste; too conservative $\rightarrow$ low benefit.~\cite{FDP} \\ \hline

\textbf{Overhead} & Extra memory bandwidth, storage, and energy consumed by prefetching.~\cite{FDP,Overhead} & Evaluates efficiency by comparing benefits vs. costs. \\ \hline


\end{tabular}
\label{tab:prefetch_metrics}
\end{table*}


\subsection{Survey Methodology}
\label{sec:methodology}

To ensure rigor, reproducibility, and comprehensive coverage, this survey follows a systematic literature review methodology aligned with established guidelines for computing surveys~\cite{kitchenham2007guidelines}. We describe the search strategy, selection criteria, screening process, and final corpus composition below.

\subsubsection{Search Strategy}
\label{sec:search-strategy}

We conducted systematic searches across five major digital libraries and indexing services between January 2024 and March 2025, with iterative updates through June 2025 to capture late-breaking publications:

\begin{enumerate}[leftmargin=*, label=(\roman*)]
    \item \textbf{IEEE Xplore} (\url{https://ieeexplore.ieee.org})
    \item \textbf{ACM Digital Library} (\url{https://dl.acm.org})
    \item \textbf{Springer Link} (\url{https://link.springer.com})
    \item \textbf{Scopus} (\url{https://www.scopus.com})
    \item \textbf{arXiv} (\url{https://arxiv.org}) --- for recent preprints not yet in indexed venues
\end{enumerate}

\noindent The search employed the following composite query string, adapted to the syntax of each database:

\begin{quote}
\small
\texttt{("data prefetch*" OR "cache prefetch*" OR "memory prefetch*" OR "hardware prefetch*") \\
AND \\
("machine learning" OR "neural network" OR "reinforcement learning" OR "deep learning" \\
OR "stride" OR "spatial" OR "temporal" OR "correlation" OR "irregular" OR "complex pattern" \\
OR "hybrid prefetch*" OR "software prefetch*" OR "compiler prefetch*")}
\end{quote}

\noindent The query was intentionally broad to capture both traditional and ML-based prefetching techniques. Additional targeted searches were performed using specific technique names (e.g., ``AMPM prefetch,'' ``Pythia reinforcement,'' ``LSTM memory access prediction,'' ``Bingo spatial prefetcher'') to ensure coverage of well-known works that might not match the general query. We also performed backward and forward citation tracking (snowballing) on highly cited foundational papers~\cite{ref1, falsafi2014primer, ref12} to capture influential works that predated our primary search window.

\subsubsection{Inclusion and Exclusion Criteria}
\label{sec:criteria}

Table~\ref{tab:inclusion-exclusion} summarizes the criteria used to determine whether a candidate paper was included in the final survey corpus.

\begin{table}[t]
\centering
\caption{Inclusion and Exclusion Criteria for Paper Selection}
\label{tab:inclusion-exclusion}
\small
\renewcommand{\arraystretch}{1.25}
\begin{tabular}{|p{0.42\columnwidth}|p{0.48\columnwidth}|}
\hline
\rowcolor[gray]{0.15}
\textcolor{white}{\textbf{Inclusion Criteria}} & \textcolor{white}{\textbf{Exclusion Criteria}} \\
\hline
Proposes, evaluates, or surveys a CPU cache prefetching technique & Focuses exclusively on GPU prefetching or GPU memory management \\
\hline
\rowcolor[gray]{0.95}
Published in a peer-reviewed venue (conference, journal, or workshop) or available as a well-cited preprint on arXiv & Non-peer-reviewed blog posts, white papers, or technical reports without evaluation \\
\hline
Addresses data or instruction prefetching at any cache hierarchy level (L1, L2, LLC, memory-side) & Addresses only cache replacement policies, coherence protocols, or TLB management without a prefetching component \\
\hline
\rowcolor[gray]{0.95}
Written in English & Focuses on network or web content prefetching (e.g., browser prefetching, CDN caching) \\
\hline
Provides sufficient technical detail (algorithm description, evaluation, or formal analysis) & Duplicate publications of the same work (earlier version excluded; most complete version retained) \\
\hline
\rowcolor[gray]{0.95}
Published between 1978 and June 2025 (foundational works included regardless of date) & Poster-only or abstract-only publications without a full paper \\
\hline
\end{tabular}
\end{table}

\noindent We note two scope decisions. First, we include storage-level prefetching (e.g., SSD and flash translation layer schemes) when the techniques are architecturally relevant to CPU cache prefetching principles, but exclude pure I/O scheduling or disk prefetching that lacks connection to cache hierarchy design. Second, we include arXiv preprints only when they present novel, technically complete contributions and have received meaningful community engagement (e.g., citations, adoption in subsequent published work, or acceptance at a recognized venue).

\subsubsection{Screening Process}
\label{sec:screening}

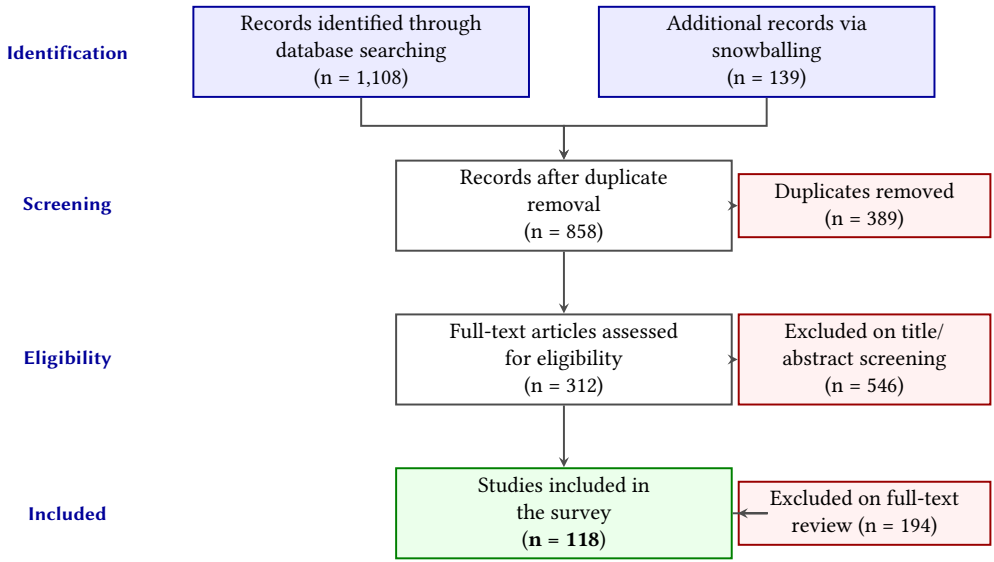
\begin{figure}[t]
\centering
\resizebox{0.95\columnwidth}{!}{%
\begin{tikzpicture}[
    box/.style={rectangle, draw=black!70, fill=white, thick, minimum width=4.8cm, minimum height=1.1cm, align=center, font=\small},
    bluebox/.style={rectangle, draw=blue!60!black, fill=blue!8, thick, minimum width=4.8cm, minimum height=1.1cm, align=center, font=\small},
    greenbox/.style={rectangle, draw=green!50!black, fill=green!8, thick, minimum width=4.8cm, minimum height=1.1cm, align=center, font=\small},
    redbox/.style={rectangle, draw=red!60!black, fill=red!5, thick, minimum width=3.6cm, minimum height=0.9cm, align=center, font=\small},
    arrow/.style={->, >=stealth, thick, draw=black!70},
    label/.style={font=\footnotesize\bfseries\sffamily, text=blue!60!black}
]

\node[label] at (-4.2, 0) {Identification};
\node[label] at (-4.2, -2.2) {Screening};
\node[label] at (-4.2, -4.4) {Eligibility};
\node[label] at (-4.2, -6.6) {Included};

\node[bluebox] (search) at (0, 0) {Records identified through\\database searching\\(n = 1,108)};
\node[bluebox] (snowball) at (5.8, 0) {Additional records via\\snowballing\\(n = 139)};

\node[box] (dedup) at (2.9, -2.2) {Records after duplicate\\removal\\(n = 858)};
\node[redbox] (excl1) at (7.2, -2.2) {Duplicates removed\\(n = 389)};

\node[box] (screen) at (2.9, -4.4) {Full-text articles assessed\\for eligibility\\(n = 312)};
\node[redbox] (excl2) at (7.2, -4.4) {Excluded on title/\\abstract screening\\(n = 546)};

\node[greenbox] (included) at (2.9, -6.6) {Studies included in\\the survey\\(\textbf{n = 118})};
\node[redbox] (excl3) at (7.2, -6.6) {Excluded on full-text\\review (n = 194)};

\draw[arrow] (search.south) -- ++(0,-0.4) -| (dedup.north);
\draw[arrow] (snowball.south) -- ++(0,-0.4) -| (dedup.north);
\draw[arrow] (dedup.south) -- (screen.north);
\draw[arrow] (screen.south) -- (included.north);

\draw[arrow] (dedup.east) -- (excl1.west);
\draw[arrow] (screen.east) -- (excl2.west);
\draw[arrow] (included.east) -- ++(0.5,0) |- (excl3.west);

\end{tikzpicture}
}%
\caption{PRISMA-style flow diagram of the systematic literature selection process. Starting from 1,247 identified records, the three-phase screening yielded a final corpus of 118 papers.}
\label{fig:prisma}
\end{figure}

The selection followed a three-phase screening process, illustrated in the PRISMA-style flow diagram in Figure~\ref{fig:prisma}:

\begin{enumerate}[leftmargin=*, label=\textbf{Phase \arabic*:}]
    \item \textbf{Identification.} The database searches and snowballing yielded a combined total of 1,247 records. After removing 389 duplicates across databases, 858 unique records remained for screening.

    \item \textbf{Screening.} Two reviewers independently screened titles and abstracts of all 858 records against the inclusion/exclusion criteria. Records clearly outside scope (e.g., web prefetching, GPU-only studies, non-English papers) were excluded, leaving 312 papers for full-text assessment. Inter-reviewer agreement at this stage was 91\%; disagreements were resolved through discussion.

    \item \textbf{Eligibility and Inclusion.} Full texts of the 312 candidate papers were reviewed for technical depth, relevance to CPU cache prefetching, and alignment with the survey's scope (Section~1). Papers lacking sufficient technical detail (e.g., position papers, 2-page abstracts) or falling outside the defined scope were excluded. This phase yielded a final corpus of \textbf{118 papers} included in the survey.
\end{enumerate}

\subsubsection{Corpus Composition and Temporal Distribution}
\label{sec:corpus-composition}

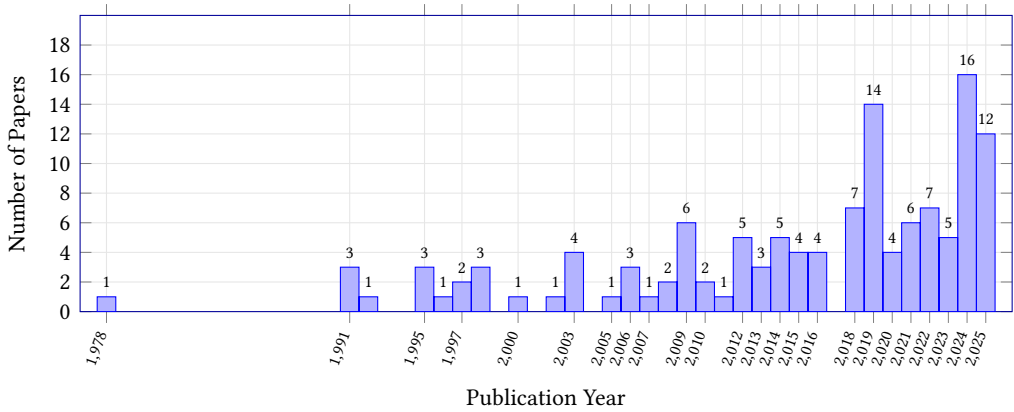
\begin{figure}[t]
\centering
\begin{tikzpicture}
\begin{axis}[
    ybar,
    bar width=7pt,
    width=\columnwidth,
    height=5.5cm,
    xlabel={Publication Year},
    ylabel={Number of Papers},
    ymin=0,
    ymax=20,
    xtick={1978, 1991, 1995, 1997, 2000, 2003, 2005, 2006, 2007, 2009, 2010, 2012, 2013, 2014, 2015, 2016, 2018, 2019, 2020, 2021, 2022, 2023, 2024, 2025},
    xticklabel style={rotate=70, anchor=east, font=\tiny},
    ytick={0, 2, 4, 6, 8, 10, 12, 14, 16, 18},
    yticklabel style={font=\small},
    ylabel style={font=\small},
    xlabel style={font=\small},
    enlarge x limits=0.03,
    grid=major,
    grid style={gray!20},
    fill=blue!45!black,
    draw=blue!60!black,
    every node near coord/.append style={font=\tiny, above, text=black},
    nodes near coords,
    nodes near coords align={vertical},
    nodes near coords style={/pgf/number format/fixed},
]
\addplot coordinates {
    (1978, 1)
    (1991, 3)
    (1992, 1)
    (1995, 3)
    (1996, 1)
    (1997, 2)
    (1998, 3)
    (2000, 1)
    (2002, 1)
    (2003, 4)
    (2005, 1)
    (2006, 3)
    (2007, 1)
    (2008, 2)
    (2009, 6)
    (2010, 2)
    (2011, 1)
    (2012, 5)
    (2013, 3)
    (2014, 5)
    (2015, 4)
    (2016, 4)
    (2018, 7)
    (2019, 14)
    (2020, 4)
    (2021, 6)
    (2022, 7)
    (2023, 5)
    (2024, 16)
    (2025, 12)
};
\end{axis}
\end{tikzpicture}
\caption{Distribution of the 118 included papers by publication year. The surge from 2018 onward reflects the growing adoption of ML-based prefetching techniques. The 2024--2025 period accounts for 23.7\% of the corpus.}
\label{fig:year-distribution}
\end{figure}

Figure~\ref{fig:year-distribution} shows the distribution of the 118 included papers by publication year. The corpus spans nearly five decades of prefetching research, from Smith's foundational work on sequential prefetching in 1978~\cite{b1} to recent ML-based and hybrid approaches published in 2025. Several temporal patterns are noteworthy.

The period 1991--2003 represents the \textit{foundational era}, during which the core spatial and temporal prefetching paradigms were established (stride prefetching, Markov predictors, correlation-based designs, and the Global History Buffer). Publication activity remained steady but modest during this period, with 1--3 papers per year in our corpus, reflecting the maturity of basic techniques.

A significant acceleration is visible from 2018 onward, coinciding with the introduction of ML-based prefetching techniques (Delta-LSTM in 2018, PPF and Bingo in 2019, Pythia in 2021). The years 2019 and 2024 represent peak publication activity in our corpus, with 14 and 16 papers respectively. This surge reflects the convergence of two trends: the growing inadequacy of traditional heuristics for irregular workloads, and the availability of mature ML frameworks that make neural and RL-based prediction feasible.

The 2024--2025 papers (28 total, representing 23.7\% of the corpus) are particularly concentrated in ML-based and hybrid approaches, confirming that the field's center of gravity is shifting toward intelligent, learning-based prefetching---the core focus of this survey.

\subsubsection{Distribution by Venue and Category}
\label{sec:venue-distribution}

Table~\ref{tab:venue-distribution} summarizes the distribution of included papers across major publication venues. The dominance of top-tier computer architecture venues (MICRO, ISCA, HPCA, ASPLOS) confirms that prefetching remains a core concern of the architecture community. The presence of systems venues (EuroSys, FAST, PPoPP) and database venues (VLDB) reflects the broadening scope of prefetching research into software systems and application-specific domains.

\begin{table}[t]
\centering
\caption{Distribution of Included Papers by Publication Venue (Top Venues)}
\label{tab:venue-distribution}
\small
\renewcommand{\arraystretch}{1.15}
\begin{tabular}{|l|c|c|}
\hline
\rowcolor[gray]{0.15}
\textcolor{white}{\textbf{Venue}} & \textcolor{white}{\textbf{Count}} & \textcolor{white}{\textbf{\% of Corpus}} \\
\hline
MICRO (IEEE/ACM Int. Symp. Microarchitecture) & 18 & 15.3\% \\
\hline
\rowcolor[gray]{0.95}
ISCA (Int. Symp. Computer Architecture) & 14 & 11.9\% \\
\hline
HPCA (IEEE Int. Symp. High-Perf. Comp. Arch.) & 11 & 9.3\% \\
\hline
\rowcolor[gray]{0.95}
ASPLOS (Arch. Support for Prog. Lang. \& OS) & 6 & 5.1\% \\
\hline
ACM Computing Surveys / IEEE TC / IEEE Micro & 8 & 6.8\% \\
\hline
\rowcolor[gray]{0.95}
arXiv (preprints with community engagement) & 12 & 10.2\% \\
\hline
Other venues (EuroSys, VLDB, MEMSYS, CF, etc.) & 49 & 41.5\% \\
\hline
\end{tabular}
\end{table}

In terms of technique categories, the corpus breaks down as follows: spatial prefetchers (26 papers, 22.0\%), temporal prefetchers (21 papers, 17.8\%), software-based prefetchers (11 papers, 9.3\%), hybrid prefetchers (15 papers, 12.7\%), ML-based prefetchers (19 papers, 16.1\%), storage-based prefetchers (3 papers, 2.5\%), and cross-cutting works covering background, metrics, benchmarking, or multiple categories (23 papers, 19.5\%). This distribution is consistent with the survey's emphasis on complex pattern prediction, where ML-based and hybrid techniques collectively represent 28.8\% of the corpus---the largest combined category.

\subsubsection{Limitations of the Methodology}
\label{sec:methodology-limitations}

We acknowledge several limitations of our search and selection process. First, our reliance on English-language publications may exclude relevant work published in other languages, particularly from the Chinese and Japanese architecture communities. Second, the inclusion of arXiv preprints introduces a quality variance, as these papers have not undergone formal peer review; we mitigated this by requiring evidence of community engagement (citations or subsequent venue acceptance). Third, our search queries may not capture every relevant paper, particularly those using unconventional terminology; we addressed this through snowballing and targeted searches for known techniques. Finally, the rapidly evolving nature of ML-based prefetching means that additional relevant works may have appeared between the completion of our search (June 2025) and the publication of this survey.

\section{Background and Terminologies}

\subsection{Prefetching Metrics}
The researchers standardize numerous metrics in order to identify the performance of prefetchers. Various well-accepted metrics are employed in the literature to systematically evaluate the effectiveness of hardware prefetching techniques. These metrics capture different aspects such as accuracy, timeliness, efficiency, and resource overhead, providing a balanced view of the strengths and limitations of the prefetcher. Table~\ref{tab:prefetch_metrics} lists the common metrics and features which are used in the evaluation of prefetchers with definitions and insights regarding their practical significance.

Higher coverage indicates that the prefetcher is able to successfully anticipate future memory accesses, thereby reducing a larger fraction of demand misses. However, coverage alone does not indicate efficiency, because a prefetcher may also generate useless prefetches, increasing bandwidth pressure.

\begin{figure}[h!]
\centering
\begin{minipage}{0.45\linewidth}
    \centering
    \includegraphics[width=\linewidth]{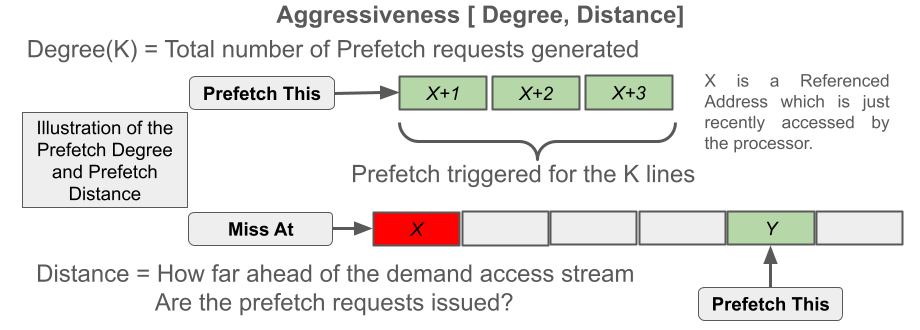}
    \caption*{(a) Visual Representation of Prefetch Degree and Prefetch Distance}
\end{minipage}\hfill
\begin{minipage}{0.45\linewidth}
    \centering
    \includegraphics[width=\linewidth]{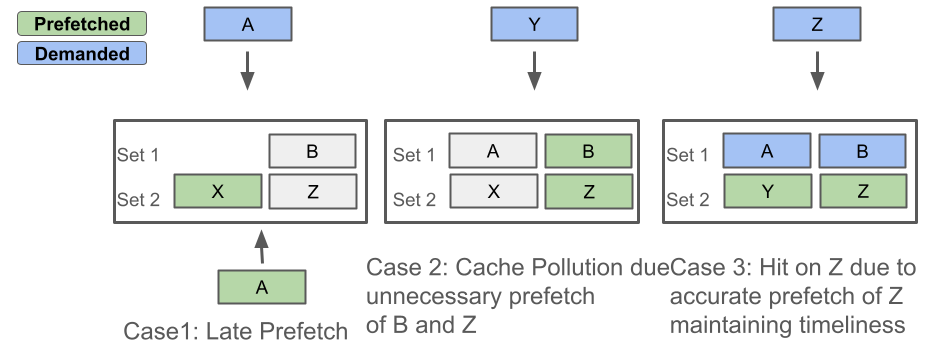}
    \caption*{(b) Case 1: Late Prefetch, Case 2: Cache Pollution, and Case 3: Timely Prefetch}
\end{minipage}
\caption{Illustration of prefetch degree and prefetch distance. (a) Shows how a demand access at address X triggers K prefetch requests ahead in the address stream. (b) Demonstrates three possible outcomes: Case 1 — late prefetch where data arrives after the demand, Case 2 — cache pollution caused by unnecessary prefetches evicting useful blocks, and Case 3 — a timely prefetch that results in a cache hit while preserving cache efficiency.}
\label{fig:prefetch_distnace}
\end{figure}

\subsection{Software and Hardware Prefetching}

Software prefetching involves inserting prefetch instructions into program code by the programmer or compiler. These instructions prompt the memory controllers to fetch data from the cache hierarchy before it is required by the currently running application. Thus, it blends memory access latencies with efficient computation; software prefetching improves overall system performance by essentially masking memory delay.  In contrast, hardware prefetching aims to increase memory access speed by forecasting future data demands and transforming prospective cache misses into cache hits.  During program execution, the CPU monitors memory consumption and, when predictable patterns such as successive array visits are detected, the relevant data is cached.  Software prefetching uses the programmer's or compiler's understanding of the application's memory access patterns for accuracy, whereas hardware prefetching uses runtime patterns to forecast future accesses.~\cite{Vanderwiel}. As a result, when the CPU requests prefetched data, it significantly reduces retrieval time compared to accessing main memory. The memory controller autonomously identifies access patterns and retrieves data or instructions anticipated for future use. This article primarily provides an in-depth exploration and analysis of various hardware-based prefetching techniques~\cite{b2,Old2,Old3,Old5}.

Before moving to the detailed discussion and classification of the hardware prefetching techniques in Section~\ref{HardwarePrefetching}. The locality of reference refers to the pattern in which programs access memory and is of two main types: temporal locality, where recently accessed data is likely to be accessed again soon, and spatial locality, where memory locations near a recently accessed address are likely to be accessed shortly. \textit{Temporal prefetching} methods take advantage of repetitive patterns in cache misses over time. When the processor encounters a cache miss, a temporal prefetcher examines past memory accesses to detect repeated sequences, called streams. 
Identification of such a stream controller can proactively load the address into the cache. This approach significantly reduces future cache misses and enhances the overall efficiency of data retrieval~\cite{ref9,ref10,ref11,ref12}. A spatial prefetcher depends on the correlation between addresses within the same memory region. \textit{Spatial prefetching} methods exploit pattern recognition over the spatial region to reduce compulsory misses. The minimal design and storage requirements of the spatial prefetcher make it deployable over the system. It uses the \textbf{offset} (\textit{the distance of a particular data block from the start of its memory page}) and the \textbf{delta} (\textit{the distance between two consecutive accesses within the same page}) for future references~\cite{b7,ref26}. 
Acccording to figure~\ref{fig:prefetch_distnace}, Prefetch Degree is the number of cache lines prefetched per request — if the processor accesses address X, a degree of 3 means X+1, X+2, and X+3 are all prefetched. Prefetch Distance is how far ahead of the current demand access the prefetch is issued; a larger distance gives the data more time to arrive before it's needed.
These two parameters together define prefetch aggressiveness. Getting them wrong leads to three outcomes: (1) Late Prefetch — the prefetched data arrives after the processor already needs it, causing a stall; (2) Cache Pollution — unnecessary prefetches evict useful data (e.g., prefetching B and Z displaces something the program actually needs); and (3) Timely Prefetch — the ideal case where the right data is fetched early enough to be in cache exactly when demanded, delivering a hit without wasting space.

\begin{figure*}[t]
\centering
\resizebox{\textwidth}{!}{%
\begin{forest}
  for tree={
    my node,
        font=\scriptsize,   
    l sep+=3pt,
    s sep=5pt,
    grow'=east,
    edge={gray, thin},
    parent anchor=east,
    child anchor=west,
    if n children=0{tier=last}{},
    edge path={
      \noexpand\path [draw, \forestoption{edge}] (!u.parent anchor) -- +(8pt,0) |- (.child anchor)\forestoption{edge label};
    },
    if={isodd(n_children())}{
      for children={
        if={equal(n,(n_children("!u")+1)/2)}{calign with current}{}
      }
    }{}
  }
  [Prefetcher
    [Hardware-Based
      [Spatial
        [Offset Based]
        [Sequential
          [Stream Based]
          [Nextline Based]
        ]
      ]
      [Temporal
        [History Based]
        [Correlation Based]
      ]
    ]
    [Software-Based
      [Compiler Based]
      [Semantic Based]
    ]
    [Hybrid Approach
      [Software-Guided Hardware]
      [Hardefdware-Assisted Software]
    ]
  ]
\end{forest}%
}
\caption{Taxonomy of prefetchers spanning both columns.}
\label{fig:prefetcher-taxonomy}
\end{figure*}
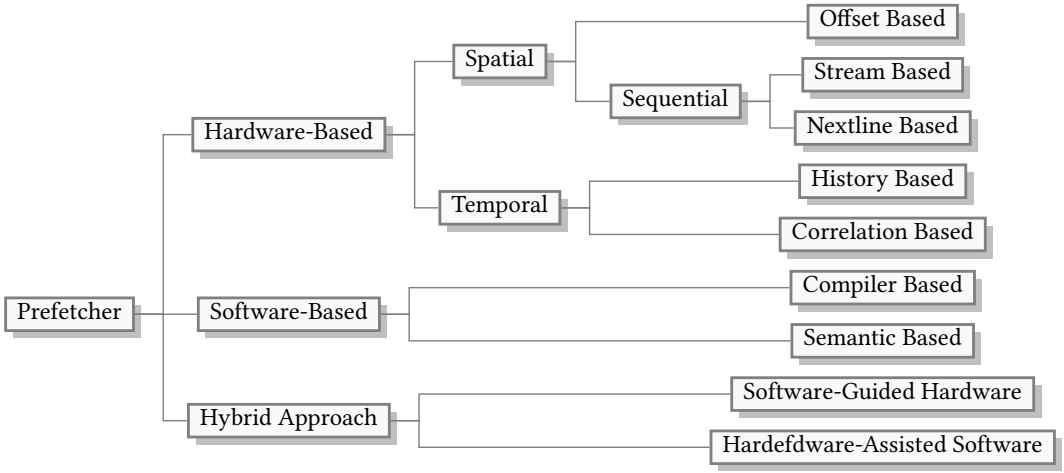

Most prefetching strategies broadly utilize temporal and spatial locality for future references. However, equal importance must be given to the type of data we are prefetching, whether it's data or an instruction. Because both highlight distinct ways of consideration, for example, instructions mostly possess sequential execution, whereas predicting data differs from the subsequent running application. This entirely depends on the program's behavior. Figure~\ref{fig:prefetcher-taxonomy} explains the taxonomy of prefetching techniques, showing hardware-, software-, and hybrid-based approaches, with further subdivisions such as spatial, temporal, compiler-guided, and cooperative designs.

Prefetch instructions before they are needed improves the performance of the instruction cache. Assuming instructions will be accessed sequentially, it prefetches the next few instructions considered as Sequential Prefetching. We can prefetch instructions from predicted branch targets, which is branch target-based prefetching. On the other hand, data-based prefetching focuses on preloading data into the cache before it is requested. Where prospective data could be prefetch in the form of strides or stream. Detects regular access patterns (e.g., arrays) and prefetchs future addresses following the same stride, whereas identifying streaming access patterns and fetches subsequent data can be categrogized under the stream based prefetching. Previous research categorizes into 1) \textbf{Sequential Prefetching}~\cite{b1}: Retrieves the subsequent contiguous block of data or instructions, supposing a linear access pattern.2) \textbf{Stride Prefetching}~\cite{b2}: Retrieves data or instructions according to a recognized access pattern characterized by a consistent stride or offset. 3) \textbf{Correlated Prefetching}~\cite{ref27,Temporal4,ref29}: Using previous access patterns and correlations to anticipate future accesses and prefetch appropriately.~\cite{ref30,Temporal8}

\tikzset{
  my node/.style={
    draw=gray,
    inner color=gray!5,
    outer color=gray!5,
    thick,
    minimum width=0.5cm,
    text height=0.6ex,
    text depth=0ex,
    font=\sffamily,
    drop shadow,
  }
}

\begin{table*}[htbp]
\centering
\footnotesize
\caption{Prior Art of prefetching techniques and its categorization. 
$\clubsuit$ $\rightarrow$ Stride, 
$\bigstar$ $\rightarrow$ Adaptive, 
$\varclubsuit$ $\rightarrow$ Delta, 
$\circledast$ $\rightarrow$ Offset, 
$\blacktriangle$ $\rightarrow$ History, 
$\spadesuit$ $\rightarrow$ RL-Based, 
$\varspadesuit$ $\rightarrow$ Signature-Based, 
$\heartsuit$ $\rightarrow$ Policy Selection-Based, 
$\blacklozenge$ $\rightarrow$ In-Memory Friendly}
\label{tab:sorting_comparison}
\begin{tabular}{|c|l|c|p{5cm}|c|}
\hline
\textbf{Category} & \textbf{Reference} & \textbf{Year} & \textbf{Technique} & \textbf{Remarks} \\
\hline

\multirow{12}{*}{Spatial}&Somogyi et al.~\cite{ref32} & 2006 & SMS (Spatial Memory Streaming) &  \\ \cline{2-5}
&Ishii et al.~\cite{AMPM} & 2009 & AMPM & $\clubsuit$ \\ \cline{2-5}
&Pugsley et al.~\cite{Sandbox} & 2014 & Sandbox prefetcher & $\bigstar$ , \\ \cline{2-5}
&Kadjo et al.~\cite{ref36} & 2014 & B-Fetch & $\clubsuit$ \\ \cline{2-5}
&Shevgoor et al.~\cite{VLDP} & 2015 & VLDP & $\clubsuit$,$\varclubsuit$ \\ \cline{2-5}
&Michaud~\cite{b9} & 2016 & BOP & $\circledast$,   \\ \cline{2-5}
&Kim et al.~\cite{ref37} & 2016 & Lookahead prefetching &  \\ \cline{2-5}
&Bera et al.~\cite{ref26} & 2019 & DSPatch &  \\ \cline{2-5}
&Bakhshalipour et al.~\cite{b7} & 2019 & Bingo & $\blacktriangle$,$\blacklozenge$,$\varspadesuit$\\ \cline{2-5}
&Shakerinava et al.~\cite{MLOP} & 2019 & MLOP & $\circledast$,  $\bigstar$ \\ \cline{2-5}
&Chen et al.~\cite{chen2024gaze} & 2024 & + temporal pattern &  \\ \cline{2-5}
&Xue et al.~\cite{ref3} & 2024 & Tyche & \\ \hline

\multirow{13}{*}{Temporal}
&Wenisch et al.~\cite{ref21} & 2002 & Temporal streaming & $\blacktriangle$ \\ \cline{2-5} 
&Hu et al.~\cite{Temporal8} & 2003 & TCP prefetcher &  $\blacktriangle$ \\ \cline{2-5} &Solihin et al.~\cite{ref29} & 2003 & Correlation prefetching &  \\ \cline{2-5} &Nesbit \& Smith~\cite{Temporal6} & 2005 & GHB & $\blacktriangle$ \\ \cline{2-5} 
&Thaker et al.~\cite{Temporal3} & 2006 & Quantum memory &  \\ \cline{2-5} 
&Srinath et al.~\cite{FDP} & 2007 & FDP & $\bigstar$,  \\ \cline{2-5} 
&Wenisch et al.~\cite{STMS} & 2009 & STMS & $\blacktriangle$ \\ \cline{2-5} 
&Wenisch et al.~\cite{ref30} & 2010 & Address correlation & $\blacktriangle$ \\ \cline{2-5} 
&Jain \& Lin~\cite{Temporal4} & 2013 & ISB & $\blacktriangle$,$\varspadesuit$ \\ \cline{2-5} 
&Wu et al.~\cite{MISB} & 2019 & MISB & $\blacktriangle$,$\varspadesuit$ \\ \cline{2-5} 
&Wu et al.~\cite{Triage} & 2019 & Triage & $\blacktriangle$,$\varspadesuit$ \\ \cline{2-5} 
&Singh et al.~\cite{ref27} & 2024 & AMC & $\bigstar$ \\ \cline{2-5} 
&Wu et al.~\cite{Temporal5} & 2025 & AdaTP & $\bigstar$ \\ \hline

\multirow{5}{*}{Software Based}
&Hummer et al.~\cite{sem2} & 2014 & Context-aware prefetching &   \\ \cline{2-5}
&Agarwal et al.~\cite{ref20} & 2015 & Adaptive software prefetching &  $\bigstar$ \\ \cline{2-5}
&Peled et al.~\cite{sem1} & 2020 & Semantic prefetching &   \\ \cline{2-5}
&APT-GET~\cite{APT-GET} & 2022 & Profile-guided prefetching &   \\ \cline{2-5}
&Kühn et al.~\cite{Roland} & 2024 & Software prefetch analysis &   \\ \hline

\multirow{8}{*}{Hybrid}
&Wang et al.~\cite{GRP} & 2003 & Guided region prefetching &  \\ \cline{2-5}
&Ebrahimi et al.~\cite{HybridPrefetch} & 2009 & Linked data prefetching &  \\ \cline{2-5}
&Ainsworth \& Jones~\cite{ProgP} & 2018 & Programmable prefetcher &  \\ \cline{2-5}
&Zhang et al.~\cite{RnR} & 2020 & Record \& replay & ,$\heartsuit$ \\ \cline{2-5}
&Alcorta et al.~\cite{ref11} & 2023 & Decision-tree runtime selection & $\bigstar$ \\ \cline{2-5}
&Li et al.~\cite{hopp} & 2023 & HoPP &  \\ \cline{2-5}
&Dong et al.~\cite{DTAP} & 2025 & DTAP & ,$\heartsuit$ \\ \cline{2-5}
&Li et al.~\cite{PGTP} & 2025 & Prophet & $\bigstar$,$\heartsuit$ \\ \hline

\multirow{15}{*}{ML Based}
&Peled et al.~\cite{ref5} & 2018 & Neural prefetching &  $\blacktriangle$ \\ \cline{2-5}
&Hashemi et al.~\cite{Delta-LSTM} & 2018 & Delta-LSTM & $\blacktriangle$,$\varclubsuit$ \\ \cline{2-5}
 &Srinivasan et al.~\cite{ref23} & 2019 & Perceptron prefetcher &  $\spadesuit$ \\ \cline{2-5}
 &Bhatia et al.~\cite{PPF} & 2019 & PPF &   \\ \cline{2-5}
 &Srivastava et al.~\cite{CompressedLSTM} & 2019 & Compressed LSTM &  $\blacktriangle$ \\ \cline{2-5}
 &Peled et al.~\cite{ref8} & 2019 & Neural prefetcher &   \\ \cline{2-5}
 &Zhang et al.~\cite{RAOP} & 2021 & RAOP &  $\blacktriangle$,$\varclubsuit$ \\ \cline{2-5} 
 &Bera et al.~\cite{pythia} & 2021 & Pythia (RL) &  $\spadesuit$, $\bigstar$,$\heartsuit$ \\ \cline{2-5}
 &Shi et al.~\cite{voyager} & 2021 & Voyager &  $\blacktriangle$ \\ \cline{2-5}
 &Zhang et al.~\cite{ref6} & 2022 & Transformer prefetching &  $\blacktriangle$ \\ \cline{2-5}
 &Bhattacharjee et al.~\cite{ref25} & 2022 & GNN prefetching &  $\spadesuit$ \\ \cline{2-5}
 &Jamet et al.~\cite{Jamet_2024} & 2024 & Two-level neural &   \\ \cline{2-5}
 &Jia et al.~\cite{b10} & 2024 & SNN-based prefetching &   \\ \cline{2-5}
 &Wang et al.~\cite{ref15} & 2024 & CNN-LSTM & $\blacktriangle$ \\ \cline{2-5}
&Siddiqui et al.~\cite{siddiqui2025} & 2025 & RL multicore prefetching & $\bigstar$,$\spadesuit$ \\ \hline

\multirow{2}{*}{Storage Based}
&Li et al.~\cite{ref16} & 2022 & SSD prefetching &  $\blacklozenge$ \\ \cline{2-5}
&Garg et al.~\cite{CrossPrefetch} & 2024 & Cross-layer I/O prefetching &  $\blacklozenge$ \\ \hline

\end{tabular}
\end{table*}
\section{Hardware Based Prefetching}\label{HardwarePrefetching}

Hardware data prefetching bridges the gap by proactively fetching data ahead of the cores’ requests to eliminate idle cycles in which the processor waits for the memory system's response. In this section, we briefly review the  approaches that target the same goal (i.e., bridging the processor-memory performance gap) but in other ways. Recent works continue to push the envelope with more sophisticated hardware prefetchers that leverage spatial and temporal correlations, low-cost neural predictors, and coordinated multi-core designs.~\cite{chen2024gaze,Jamet_2024,siddiqui2025,ref34,ref1,ref35}

\subsection{Spatial Prefecthers}

Spatial prefetchers take advantage of the spatial locality in memory accesses, for example, contiguous cache lines within arrays, and clustered addresses in structured data. In the spirit of coverage maximization, a spatial prefetcher anticipates memory blocks that are adjacent to the most recently served demand request. Traditional methods like next-line and sequential prefetching put this concept into practice in a certain way. Often during irregular workloads this proposal led to overprediction or over-fetching, impacting overall coverage. To resolve this issue many designs have adaptive mechanisms which learn the pattern of accessing the memory, observe the spatial correlation patterns between the memory regions, and selectively prefetch them according to the observed program behavior. Employing an adaptive approach not only decreases the bandwidth waste but also guarantees high coverage~\cite{FDP,chen2024gaze}.

\subsubsection{\textbf{Stream Based: Prefetching on the basis of fixed stream}}

\begin{figure}
    \centering
    \includegraphics[width=0.5\linewidth]{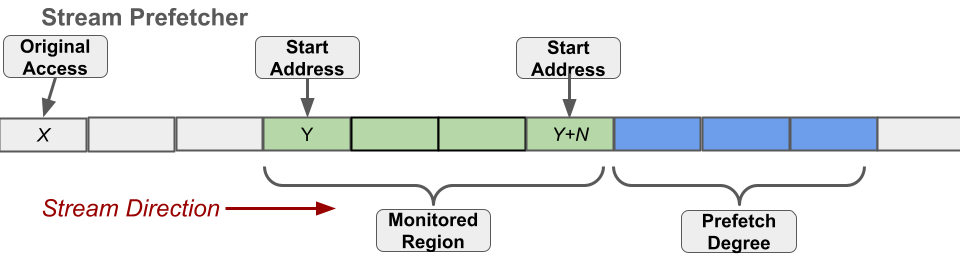}
    \caption{Illustration of Generaic Stream Based Prefetcher.}
    \label{fig:streamPrefetch}
\end{figure}

A fixed stream-based prefetcher (Visual representation in figure:~\ref{fig:streamPrefetch}) is a fundamental hardware prefetching mechanism that operates on the assumption of sequential memory access patterns. It observes the addresses accessed by the processor and looks for a constant stride between successive accesses – typically \textbf{a stride of one cache block, corresponding to purely sequential access}. Once such a stream is detected (for example, if the system sees a memory access to block X followed by block X+1), the prefetcher will automatically fetch the subsequent blocks in that sequence in advance. The stride and the number of blocks to prefetch ahead (often called the prefetch depth or degree) are fixed design parameters. In other words, this prefetcher is non-adaptive and always assumes the same stride length (commonly +1) and will fetch a predetermined number of subsequent blocks. This simple strategy leverages the typical case of high spatial locality in programs. Fewer decisions need to be made and less complicated hardware is an advantage of such rules. Applications that access data sequentially, such iterating over array members in memory order, benefit greatly from this sort of prefetcher in terms of speed. In such instances, the necessary data is usually already in the cache when the CPU needs it. Such prefetchers give a noticeable speed boost if instruction prefetching is available.

The deterministic nature of prediction makes the prefetcher easy to implement. However, the fixed nature of this prefetcher also introduces limitations that must be addressed. As it does not adjust to actual program access patterns beyond the initial detection of a sequential stream, it can mispredict the future access stream if the program's behavior deviates from simple stride-based access. For instance, if a program accesses blocks X and X+1 (triggering the prefetcher to fetch X+2, X+3, etc.), but then the program's next access is not X+2, those prefetched blocks are unnecessary. Therefore, an \textbf{overprediction} by such prefetchers wastes memory bandwidth~\cite{ref39} and may fill the cache with irrelevant data, a phenomenon known as \textbf{cache pollution}~\cite{ref38}. A low precision in the prefetcher guesses can degrade performance rather than improve it, as useless prefetched blocks evict valid data and consume resources. Additionally, the fixed prefetch depth presents a trade-off, i.e., increasing the number of blocks prefetched ahead can improve coverage (capturing longer streams of relevant data), but will proportionally increase the risk of fetching unused blocks if the sequential pattern does not continue as expected. This tradeoff means that an \textbf{aggressive fixed stream prefetcher} might prefetch many blocks to cover potential future accesses, yet it might overshoot and fetch a stream that the program never actually follows. Considering the above limitations and tradeoffs, we have compiled a few improvisations on the stream-based prefetchers. 

A fixed stream-based prefetcher detects simple sequential access patterns and prefetches subsequent blocks based on \textbf{a constant stride(Stride)} and fixed distance from the base address. Its advantages lie in its simplicity and effectiveness for workloads with predictable linear access patterns, where it can substantially hide memory latency. However, its \textbf{lack of adaptiveness} can lead to precision issues - incorrect forecasting of streams - necessitating researchers' careful consideration. Designers must address potential issues like cache pollution and wasted bandwidth by conservatively tuning the prefetch stream and depth by incorporating safeguards. In particular, numerous academics focused on unique prediction strategies to increase the \textbf{rationality of stream prefetchers}.

\subsubsection{\textbf{Offset based prefetching}}

Offset-based prefetching represents a hardware technique that identifies regular memory access patterns characterized by a constant offset between successive address requests from the processor. These offsets between the address requests could be dynamic and arbitrary in nature. Specifically, we will calculate the correct offset at runtime by carefully tracking the delays between consecutive memory accesses associated with the fetched instructions. Once a stable offset is obtained, future prediction is taken using a stable stride calculated from an extrapolated current address. This algorithm can move forward and backward through memory, enabling a variety of patterns to be captured, which is something branched applications often employ. The general data layout has positive and negative strides~\cite{Old4}. 

Calculating stride is dynamic in nature, so while this approach works, the offset has varying ranges from use to use. There are some offset values that enable us to effectively apply a specific set of offsets at runtime. In the case of multiple applications running across different cores, the central issue is that we have to apply an effective offset value to prefetch the block. We can serve requests from the processor for irregular memory access if we can handle a decision about when to use a particular stride. A dynamic approach to offset calculation does not inherently guarantee a reduction in miss rate. While accurate offset detection can yield precise prefetches that lower both miss rates and memory stall times, the effectiveness of offset-based prefetching remains highly sensitive to the fidelity of that detection. As access patterns shift at runtime, the risk of false offset identification increases, potentially triggering unnecessary prefetch requests that introduce cache pollution rather than alleviate it.

Multiple offsets are needed for irregular memory accesses, and thus require history tracking. Then we save numerous strides in a buffer to give agility during prefetching. Multiple methods have been proposed to recognize stride-based access behavior. One possible approach is to analyze the most recent accesses and predict the offset. If it is a new access, build the address history. For instance, consider building a cumulative judgment using several offsets to keep track of previous addresses. Per-load stride histories are another limitation due to buffer capacity constraints. This adds another field for the prefetcher study. In Per-Page History (PPH) each memory page has a unique access pattern. A second method, referred to as Shared History, records patterns of access over a number of pages. These simple, yet effective, methods make cache performance and flexibility better. And regardless of those limitations, stride-based prefetching is one of the most basic techniques in cache hierarchy — to balance simplicity with flexibility and performance on the variety of frequent access workloads.~\cite{Old4}

It is worth noting that we do not provide a detailed discussion of the \textbf{next-line prefetcher}, as it represents the most trivial form of spatial prefetching and has been extensively studied in prior literature. Its simplicity, while historically important, offers limited insight into the advances that drive modern designs. Moreover, we also avoid treating adaptive prefetchers as a standalone category, since their defining mechanisms, such as dynamic aggressiveness control, footprint tracking, and feedback-based refinement, often appear as enhancements within spatial prefetchers themselves. In practice, adaptivity(\textbf{Adaptive Prefetcher}) serves as a feature to increase the rational decision-making ability of prefetchers, particularly for handling irregular and complex memory access patterns, rather than forming an entirely distinct class of techniques.

\subsubsection{\textbf{Work based on Spatial Prefetchers}}
From fundamental next-line and stride-based designs to sophisticated processes that take use of intricate spatial correlations among memory areas, research on spatial prefetchers has progressed. Due to their limited flexibility, early methods that concentrated on fixed or sequential streams frequently resulted in cache pollution and bandwidth waste in workloads that were irregular. More advanced techniques that monitor fine-grained memory footprints, examine spatial access maps, and dynamically modify aggression in response to run-time feedback were proposed in later work. These developments made it possible for spatial prefetchers to increase accuracy and efficiency while maintaining high coverage, which resulted in ground-breaking ideas like Spatial Memory Streaming (SMS).~\cite{ref32}, Access Map Pattern Matching (AMPM)~\cite{AMPM}, and Feedback-Directed Prefetching (FDP)~\cite{FDP}. The following discussion summarizes these works and highlights how each technique addresses trade-offs between accuracy, timeliness, and hardware overhead in spatial prefetching.

The paper Spatial Memory Streaming (SMS) addresses inefficiencies in memory systems due to complex data access patterns in commercial applications. SMS identifies spatial correlation in access patterns linked to specific code instructions and predicts future data requests, fetching them ahead of need. This approach significantly reduces memory errors (58\% in primary caches, 65\% in off-chip caches), improving performance by approximately 37\%. It incurs an overhead of small hardware structures integrated into existing caches hierarchy~\cite{ref32}.

Access Map Pattern Matching (AMPM)~\cite{AMPM} is a stream-based prefetcher that monitors multiple streams within a physical page and compares previously and newly prefetched addresses. Technically addressing the limitations of the traditional stride-based prefetchers, which often fail when memory accesses are reordered by compiler optimizations. The compiler can rearrange instructions, which may fetch an instruction, allowing out-of-order execution. AMPM maintains a bitmap structure, known as a memory access map, to represent the accessed cache lines within a memory region. Hardware-based pattern matching logic identifies consistent memory access patterns from this bitmap, independent of the temporal sequence of accesses. This feature allows AMPM to detect stride patterns even under significant instruction and access reordering. These methods are suitable for non-uniform cache architectures (NUCA), whereas the model training time is a main issue in this proposal. 

The AMPM prefetcher differs from the SMS prefetcher primarily in two aspects. First, SMS relies on instruction addresses to trigger prefetching, which negatively impacts its performance under optimizations like loop unrolling. Second, SMS requires a large, off-chip pattern history table, increasing complexity. In contrast, AMPM operates without relying on instruction addresses and maintains a significantly smaller, cost-effective pattern table (~5.2 KB), entirely feasible for implementation of on-chip memory. Although the spatial memory streaming method with rotated patterns attempts to reduce SMS table size, it did not achieve comparable performance.

  In Feedback-Directed Prefetching (FDP), Srinath~\cite{FDP} examined various attributes to calibrate the model's aggressiveness. We can control aggressiveness using AMPM, whereas analyzing the repercussions on performance and bandwidth is done by FDP. In AMPM, performance degradation is considered a factor in regulating the degree of prefetch. FDP regulates the degree to prefetch, assessing precision, delay, and cache contamination, which makes the prefetcher adaptive in nature, depending on performance.  Following comprehensive input, the stream was rectified for all apps, which is a restriction of this methodology. The mechanism requires maintaining multiple hardware counters (for accuracy, lateness, pollution), a “pollution” tracking filter, and adaptive logic to dynamically adjust prefetch aggressiveness. Even this model ensures cache pollution, since it has a Bloom filter structure for approximate cache pollution. This adds non-trivial complexity to the memory subsystem that leads to overhead. Identifying which demand misses due to prefetch pollution should ideally consist of tracking every displaced block, which is an impractical hardware burden. A Bloom filter gives a means of estimating, trading off accuracy for complexity. Essentially, adaptive prefetching can boost performance, but it adds to both hardware complexity and management overhead and in some cases consumes additional storage and performance cycles.

\begin{table*}[ht]
\centering
\scriptsize
\begin{tabular}{|p{1.3cm}|p{3.5cm}|p{3.5cm}|p{3.5cm}|}
\hline
\textbf{Aspect} & \textbf{Fixed Stream-based} & \textbf{Offset/Stride-based} & \textbf{Next-line based} \\
\hline
Stream Type & Sequential accesses & Detects arbitrary constant stride patterns between accesses & Assumes next sequential block (+1) after every demand fetch \\
\hline
Adaptivity & No & Yes, dynamically learns stride at runtime & No, purely reactive and fixed behavior \\
\hline
Cost & Very low & Moderate (stride detection tables requires, confidence counters) & simple hardware logic \\
\hline
Pros & Linear Memory access & Regular, non-unit stride patterns; array traversal with fixed step sizes; loop tiling & Sequential or spatially local workloads with tight loops \\
\hline
Cons & Random or irregular accesses in the applications & Irregular, data-dependent, or nonlinear access patterns & Irregular, random, or pointer-based access patterns; may cause cache pollution \\
\hline

\end{tabular}
\caption{Comparison of Fixed Stream-based, Offset/Stride-based, and Next-line Prefetchers}
\label{tab:prefetcher_comparison}
\end{table*}

The Sandbox Prefetcher~\cite{Sandbox} reduces complexity and storage overhead compared to FDP~\cite{FDP} and AMPM. Prefetchers can be categorized as conservative or aggressive~\cite{ref33}. There are conservative confirmation-based prefetchers like \textbf{stream prefetchers} that wait to notice predictable access patterns before issuing prefetches in order to prioritize accuracy over coverage. Instead, aggressive prefetchers like \textbf{next-line prefetchers}, which prefetch the next block immediately after each access, try to reach a larger coverage with little verification. This is a trade-off between avoiding unnecessary prefetching and optimizing timeliness. The sandbox prefetcher~\cite{Sandbox} strives to strike a balance by testing multiple candidate prefetchers in isolation without making any real overhead in cache or memory bandwidth. Real prefetches can only be issued to candidates with very high accuracy in predicting useful data. Each candidate aims to achieve a desired cache line offset and is tested continuously. The Sandbox Prefetcher (SBP) requires much less storage compared to FDP or AMPM, and therefore, better hardware efficiency. Its logic complexity is also smaller than that of AMPM, as it has less overhead of implementation and impact on the critical execution path. SBP is thus an increasingly well-suited fit to resource-limited settings, where the emphasis is placed on simplicity and low latency rather than on predicting aggressive patterns. Lator on researchers proposed policy to which can be used to enhance the efficiency of the Sandbox prefetcher.
\begin{figure}
    \centering
    \includegraphics[width=0.5\linewidth]{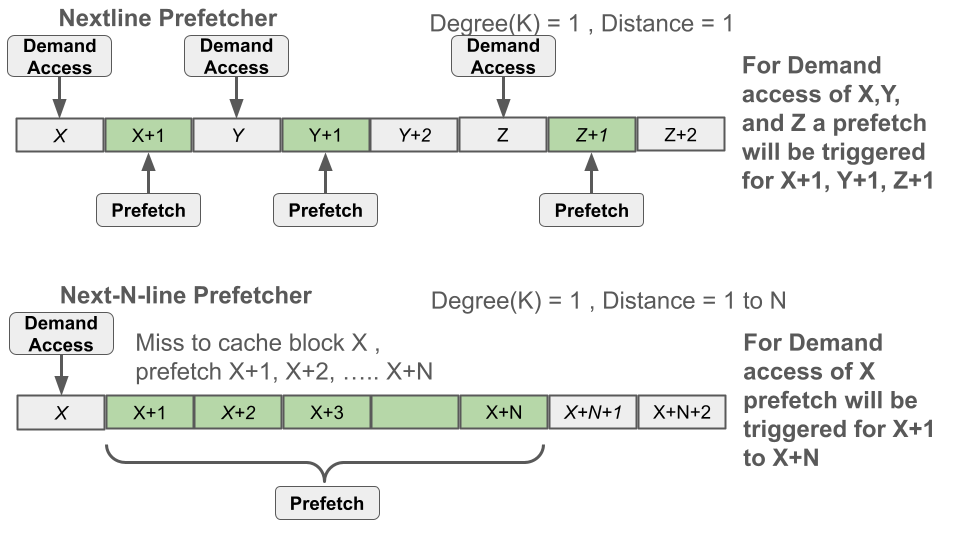}
    \caption{Illustration of Nextline and Next-N-Line Prefetcher.}
    \label{fig:nextline}
\end{figure}
We have Best Offset Prefetcher (BOP)~\cite{b9}, which enhanced the efficiency of SBP. Once we have multiple offset values in the buffer, speculating an efficient offset is a challenge. BOP ensures by predicting an approximate offset value for prefetching. For this, BOP keeps a current request database including 256 entries, each associated with a designated score for its respective offset value. The score changes in every round of the learning phase. During prefetching, the optimal offset is determined based on the score. The extent of this prefetcher is a constraint of this approach. The Shakerinava~\cite{MLOP} is addressing a similar problem of timing. This suggestion identifies cache misses as a critical factor for performance enhancement. One should use access pattern-based prefetching to enhance coverage and precision.


\subsection{Temporal Prefetchers}
A temporal cache prefetcher predicts what comes next in time rather than what comes next in space. Unlike spatial prefetchers, which rely on physical adjacency (e.g., the next cache line), temporal prefetchers exploit recurring sequences of memory accesses, even when those addresses are far apart. This makes them especially effective for irregular workloads such as pointer chasing, graph traversals, and sparse matrix operations, where spatial locality is weak but temporal correlations are strong. The issue comes when we have to deal with the memory overhead, few temporal prefetchers leverage temporal locality to prefetch cache lines but require high storage in order to maintain long memory access sequences.~\cite{ref9,MISB,Triage,voyager}

\subsubsection{\textbf{History Based Prefetchers}}
History-based prefetchers represent one of the earliest and most widely studied forms of temporal prefetching. These techniques exploit the observation that memory accesses often repeat in similar temporal sequences across program phases. By recording past miss streams in correlation tables or history buffers and using them to predict future accesses, history-based prefetchers capture long-range dependencies that spatial or stride prefetchers typically miss. Early designs such as the Markov prefetcher demonstrated the effectiveness of this approach in irregular workloads like pointer chasing and graph traversals, while later refinements improved accuracy and reduced metadata overhead~\cite{Temporal1,Temporal2,Temporal3}.

\subsubsection{\textbf{Correlation Based Prefetchers}}
Correlation-based prefetchers extend the history-based approach by exploiting relationships between memory accesses observed in the past, rather than relying solely on sequential repetition. These designs build correlation tables that record which cache misses tend to follow one another and use this information to predict future accesses. Unlike simple stride or spatial methods, correlation prefetchers can capture long-range, irregular patterns such as pointer chasing or graph traversals, making them well-suited for modern irregular workloads. 
Joseph and Grunwald introduced the idea of the correlation of addresses in the prefetcher~\cite{Temporal7}. The table maintained by the Markov predictor was larger to maintain over the chip which was optimized by the TCP~\cite{Temporal8}. The work by Nesbit and Smith~\cite{Temporal6} introduced the Global History Buffer (GHB), a highly influential design that compactly records miss sequences and enables correlation-based lookups. Later refinements, such as Delta Correlation Prefetching~\cite{Temporal4}, demonstrated how recording deltas between addresses can further improve prediction accuracy while reducing metadata overhead~\cite{Temporal5}. In order to reduce the on-chip metadata overhead researchers included localizing Program counter with the context. Program Counter (PC) localization in a temporal prefetcher is a technique that enhances prefetching accuracy and timeliness by associating memory access patterns with the specific instruction (identified by its PC) that initiated the access. This approach helps to create more predictable reference streams by segmenting them according to the source instruction. Although there is a weaker form of correlation which researchers use, such as delta correlation~\cite{Temporal6}, tag correlation~\cite{Temporal8}, and context-address pair correlation~\cite{b8}, they have limited scope. 

\subsubsection{\textbf{Work based on Temporal Prefetchers}}
The SGI Origin 2000 is a cache-coherent non-uniform memory access (ccNUMA) multiprocessor that resolves the NUMA memory latency by utilizing page migration and replication instead of conventional prefetching~\cite{Temporal1}. These approaches take advantage of temporal locality to achieve the most efficient memory access in high-throughput parallel systems. The system presumes that many pages that get accessed from one processor node will frequently be accessed again from the same node. Together, the hardware and software mechanisms work effectively together to achieve this goal. Directly in directory memory and indexed by node ID, the per-page hardware memory reference counters serve as a reference counter for maximum 64 nodes; configurations over 64 nodes share one counter among eight nodes. When a memory request is received, the reference counter for a node is incremented during directory lookup. The system compares the requester count to the home node count. If the difference is larger than the software-programmable threshold and the page is considered a migration candidate, an interrupt indicates the OS that it might be an option for migration. During migration, TLB shootdown and page copying are executed concurrently using directory poisoning. The directory goes POISON when the block copies. If the processor gets to work with a poisoned directory entry, it generates a synchronous bus error which immediately throws away the entry for the TLB and no longer requires a global TLB shootdown. This parallelization reduces the migration overhead from about 100 microseconds typical for traditional shootdown, down to almost the time for copying memory, which results in a more aggressive page migration. A dedicated block transfer engine transports 16 KB pages between nodes at near-peak memory speeds, allowing for transfers within less than 30 microseconds and a low cost migration. Temporal locality exploitation translates into altered access patterns: when a page is accessed repeatedly from a specific node, the system shifts access into that node’s local memory, converting high-latency remote accesses to low-latency local accesses. In the Origin, remote memory latency varies from 540 ns (four processors) to 945 ns (128 processors) versus 310 ns for the local latency. This costs a penalty of 1.7x to 3x, which page migration mitigates. Unlike snoopy bus methods, the directory-based cache coherence protocol keeps reference data at its per-page along with metadata about all directories provided. This allows you to make decisions over migration without having to go further and check the memory again. Since directory operations go ahead at the same time as access to memory data, managing temporal locality is an efficient enhancement to coherence operations.

The Markov Prefetcher~\cite{Temporal7} is a hardware-based memory prefetching mechanism that leverages temporal locality at the sequence level by learning and predicting patterns in the miss address stream. Operating at the interface between on-chip and off-chip cache hierarchies, it is well-suited for integration into existing computer architectures. The prefetcher models memory access patterns using a Markov chain, where each node corresponds to a cache-miss address and weighted edges represent the transition probabilities between consecutive misses. Its core mechanism maintains a prediction table indexed by miss addresses, with each entry storing up to four predicted successor addresses derived from historical observations. When a cache miss matches a table entry, all associated predictions are issued as prefetch candidates. To enhance coverage, the prefetcher triggers multiple prefetch requests. The Markov Prefetcher uses least-recently used (LRU) prioritization of predicted addresses, ensuring that the most recently learned transitions are fetched first. This approach exploits temporal locality by prioritizing recently observed patterns, thereby increasing the timeliness before processor demand. The prefetcher achieves an approximate 54\% average reduction in CPI across a range of benchmarks. It outperforms stride prefetchers and stream buffers by accurately predicting irregular, non-linear access patterns. LRU prioritization further ensures that temporally relevant data is prefetched with optimal timeliness. In contrast to stride-based prefetchers, which are limited to linear patterns, and stream buffers, which are restricted to sequential accesses, the Markov approach effectively handles irregular and unstructured memory access patterns frequently observed in commercial workloads. This system exploits temporal locality without requiring explicit programmer hints or compiler feedback. The primary innovation lies in recognizing that temporal locality exists at multiple granularities. While traditional caches leverage fine-grained temporal reuse, where the same address is accessed repeatedly, the Markov Prefetcher capitalizes on coarse-grained temporal locality present in access sequences.

The Tag Correlating Prefetcher (TCP)~\cite{Temporal8} is an implementation of hardware-efficient prefetching; it exploits temporal locality of cache tag sequences instead of full addresses. TCP has achieved a 14\% performance improvement over a baseline system, with an extra memory footprint of just 8 KB, much lower than the large number of megabytes necessary to obtain from existing address-based correlating prefetchers. The design relies on the observation that the repeat order of cache tag sequences within cache sets is predictable over time. An empirical investigation of the SPEC2000 benchmarks suggests that only 5\% of the theoretically possible three-tag sequences are actually present, implying a significant temporal relationship. For example, if a program used, say T1 to T2 to T3 the tag sequence was followed by the same subsequent tag the following time, it is probable that this sequence will recur. This repeatability allows for accurate prefetching without having to record the address in the first instance. TCP is based on two layers of architecture. For starters, the `Tag History Table` (THT) records a unique per-cache-set history of the k current miss tags (usually, k=2). As tags are implicitly indexed by the cache set, the THT performs without prediction overhead and well captures local temporal patterns. Second, the Pattern History Table (PHT) keeps track of learned tag correlation patterns globally in all cache sets. When the tag sequence happens, the PHT records the tag that usually follows. Most importantly, a single tag sequence will repeat across multiple cache sets (the average from this is 264 sets/sequence in swim benchmark) and any one PHT entry will respond to a number of address sequences. And this feature is the largest efficiency advantage over address-based designs. TCP prefetchers take advantage of the relatively high temporal recurrence that commonly characterizes real workloads, where an individual cache tag is capable of recurring at least thousands to millions of times. The art benchmark reports 3 million recurrences per tag, for instance. In addition, three-tag sequences are repeated 10–200,000 times depending on the benchmark to learn temporal patterns reliably—lessens dependence on spatial or stride-based data points. This recurrence is not limited to a single cache set; cross-set reuse is common, thus allowing tag sequences learned in a previous cache to generalize to further cache sets and contribute to the utility of prediction. TCP works as an intermediate between the L1 and L2 caches. It analyzes the L1 miss streams and issues prefetches into L2 by indexing the THT to cache set of current miss, comparing the current miss tag with the recent tag history, predicting the next tag in case of match using PHT, and prefetching the predicted address into L2 before the processor asks.

The Global History Buffer (GHB)~\cite{Temporal6} prefetcher exploits temporal locality to identify predictable repetitions in cache miss sequences. This framework stores all global miss addresses in FIFO order, and it can thus recognize and forecast temporal patterns along the miss address stream that conventional table-based methods may overlook. The GHB provides a complete, time-ordered sequence of L2 cache misses in FIFO fashion. On each cache miss, the address is added to the end of the circular buffer and removed from the front as new misses occur, thus preserving a sliding window of recent miss history and maintaining temporal relationships between consecutive misses. Unlike traditional prefetch tables which allocate a fixed-size history per entry, the FIFO format of GHB necessarily accounts for recent events. GHB also uses temporal locality more efficiently than historical methods which depend on old data by focusing on the most recent miss sequences. The GHB uses linked lists indexed by several keys (program counter or global address) to generate time-ordered sequences of misses with certain properties. The GHB achieves a 20\% improvement of Instructions Per Cycle (IPC) compared with standard distance prefetching, and a 90\% reduction in memory traffic according to empirical results. This large improvement is attributed to concentrating on temporal patterns which are quite recent and important. In addition, by maintaining a holistic temporal history, hybrid prefetching with GHB achieves 23\% higher coverage compared to simple width-based approaches. Depth and hybrid prefetching schemes are available to achieve longer look-ahead; traversing temporal chains further into predicted miss sequences ensures data is prefetched with sufficient lead time before being demanded by the processor.

Traditional table-based prefetchers with a temporal overhead (e.g. Markov prefetching), meanwhile, store a fixed history for each entry, which results in temporal decay: old patterns persist in the table even after the program performs a different behavior. Since the GHB is a FIFO, it tends to naturally weight the recent temporal patterns higher, while eliminating truly stale history, and retains new, usable temporal sequences. This guarantees that prefetches are based on currently relevant temporal patterns rather than outdated historical data.

Quantum Memory Hierarchies provides a novel memory architecture for quantum computers based on temporal locality and detecting qubit reuse patterns. The architecture separates frequently accessed qubits – held in Level-1 Memory (Compute Region) to obtain reduced latency – from infrequently accessed qubits, allocated to Level-2 Memory (Storage Region) to improve reliability although at the cost of higher latency. We can dynamically move qubits based on observed access patterns making it much easier to manage and minimizing overhead. Unlike the conventional homogeneous methods that necessitate the uniform proximity of all qubits, in this architecture space is allocated according to patterns of temporal reuse, which not only reduces physical area requirements and sustains performance, but also guarantees high performance. The advantages include reduced area of qubits, performance preservation for the popular qubits and improved reliability in Level-2 storage through strong error correction. It is shown in this work that quantum computing architectures become more feasible and scalable after following principles of temporal locality (different from classical prediction techniques).

The pros and cons of previous approaches, such as Markov prefetchers and global history buffer (GHB)-based techniques include maintaining the program counter (PC) localization, which restricts address correlation, and incorporating correlation, which reduces the benefits of PC localization. The disadvantages of the former approaches can be alleviated with structural linearization by the ISB in the form of partitioning temporal streams from different PCs while the correlations remain intact. This enables an efficient sequential prediction in the presence of irregular correlations and can even avoid resource-intensive table traversal. By preserving a temporal ordering, the ISB enhances the use of temporal locality, redefining the prediction task in a way that correlated address sequences can instead be handled by basic sequential prefetching algorithms rather than complex correlation tables.

SPMD Divergence Management on Data-Parallel Architectures~\cite{Temporal2}  focuses on the architectural trade-offs that need to be faced to control flow divergence while designing data-parallel GPU architectures. It shows that using temporal locality and how well threads could reconverge following the divergence of a thread, to get the best trade-off between the threads execution sequences and memory access pattern, may be useful to promote performance. The focus of our analysis is specifically on Hardware Divergence Stack, that is an explicit stack to manage thread paths and thus enforce temporal locality and Software Predication, a predicate-level implementation that does rely on these techniques but can sacrifice the temporal locality through interleaved memory access patterns. The results show that appropriate divergence management substantially improves memory bandwidth usage, branch predictor accuracy, and L1 cache efficiency. Using divergent strategies for temporal thread grouping improves the architecture through temporal correlation in memory accesses to achieve system performance improvements. The results suggested that the temporal locality principles can also be applied not just to prefetching but also to thread execution arrangements in GPU architectures.

STMS~\cite{STMS} is an address-correlating prefetcher that is designed to optimize temporal locality by predicting sequences of correlated cache misses while significantly reducing on-chip storage requirements. This system requires only 2 KB of on-chip storage and 64 to 96 MB off-chip storage. STMS also contributes to enhancing temporal prefetching with a hash-based index that optimizes access to off-chip cache pages, also separating it from a history buffer, enabling multiple prefetches per lookup. Update bandwidth is minimized thanks to probabilistic recording methods that limit the count of update packets while maintaining coverage. The system recognizes stream boundaries to allow for efficient prefetching, leading to an average L2 miss coverage of 56\%; in some workloads, 40–60\% of off-chip misses are successfully avoided. STMS has a small storage footprint compared to previous methods, proving effective for high-performance scenarios. Previously, temporal prefetchers such as Markov, GHB, and TCP needed megabytes of on-chip storage due to correlation tables scaling with the underlying data set of the application, usually exceeding 64 MB. Doing so rendered them impractical for real systems. Moving meta-data off-chip presented additional hurdles, such as high lookup latency, higher memory bandwidth stress due to meta-data accesses, and difficulties in amortizing lookup costs.

Domino~\cite{ref9} is a temporal data prefetcher that leverages temporal locality by identifying predictable repetitions in cache-miss sequences within server workloads. Its primary innovation lies in utilizing both single and paired consecutive miss addresses for history lookup, resulting in a 16-fold speedup over baseline systems and a sixfold improvement compared to advanced temporal prefetchers such as STMS~\cite{STMS} (Sampled Temporal Memory Streaming). Domino captures over 90\% of the theoretical temporal prefetching potential. Conventional temporal prefetchers, including STMS, rely on single-address lookups to identify streams in history, which often leads to incorrect stream selection when multiple streams share the same initial address. This limitation results in shorter predicted streams and reduced accuracy. In contrast, two-address lookups, as implemented in approaches like Digram, can identify longer and more accurate streams but may forgo prefetch opportunities by postponing the initial prefetch. Domino addresses these challenges by integrating both one- and two-address lookups. The initial lookup employs a single miss address to prefetch the subsequent address immediately. Upon the next miss or prefetch hit, two consecutive miss addresses are used to identify the correct extended stream. A history table, implemented as a circular buffer, records the time-ordered sequence of cache misses, thereby preserving the temporal order necessary for precise pattern prediction.

\section{Software Based Prefetching}
Software-based prefetching is a widely explored technique that relies on compiler- or programmer-inserted instructions to predict and prefetch data into the cache. Unlike hardware-based approaches, which rely on dedicated prefetching logic at runtime, software prefetching utilizes static or profile-guided analysis to identify predictable access patterns, such as loops or pointer traversals. It either inserts or triggers prefetch instructions according to the required optimization. This approach offers the advantage of prefetching according to the specific characteristics of the workload. It reduces unnecessary traffic on the bus and improves overall cache utilization. Notable studies~\cite{ref19,ref20} on compiler-based prefetching have demonstrated the effectiveness of software-based schemes in a range of complex access patterns. In the following section, we highlight representative examples of software-based prefetching techniques and their applications. The broad classification is in terms of Compiler based prefetching, and semantic-based prefetching.

\begin{table*}[h]
\centering
\scriptsize
\begin{tabular}{|p{2cm}|p{4cm}|p{6.8cm}|}
\hline
\textbf{Category} & \textbf{Type} & \textbf{Description / Example} \\ \hline

\textbf{Compiler-Based }& Loop-based prefetching & Inserts prefetches into regular loops (e.g., irregukar array traversals)~\cite{ref19} \\ \cline{2-3}
& Pointer-based prefetching & Targets irregular patterns in linked lists, trees, graphs~\cite{ref20} \\ \cline{2-3}
& Profile-guided prefetching & Uses profiling data to decide placement and timing~\cite{APT-GET} \\ \cline{2-3}
& Adaptive compiler prefetching & Adjusts distance and placement dynamically based on workload~\cite{ref20} \\ \hline

\textbf{Semantic-Based } 
& Type-aware prefetching & Leverages data type metadata (e.g., DTAP system)~\cite{DTAP} \\ \cline{2-3}
& Vector semantics-based prefetching & Uses vector programming semantics (e.g., ReVeLA with RVL-GVL)~\cite{ReVeLA} \\ \cline{2-3}
& Program annotation-guided prefetching & Relies on programmer-supplied hints or code annotations~\cite{RnR} \\ \hline

\end{tabular}
\caption{Summary of compiler-based and semantic-based prefetching categories}
\label{tab:prefetch_summary}
\end{table*}

 \subsection{Compiler Based Prefetchers}
APT-GET~\cite{APT-GET} is a profile-guided dynamic software prefetching technique developed to overcome the limitations inherent in static prefetchers. APT-GET triggers prefetch by analysing information gathered during program execution. This process enhances prefetching accuracy and timeliness by predicting memory accesses in advance. They have reported that modern applications can lose over 60\% of processor cycles due to cache errors; efficient prefetching can reduce the memory wall problem. APT-GET utilizes Intel's Last Branch Record (LBR)~\cite{intel_sdm_vol3_lbr} for profiling; thus, it collects execution data with significant overhead. This data helps to determine optimal prefetch distances and the most suitable locations to insert prefetch instructions. APT-GET also decides whether to add prefetch instructions to outer or inner loops. It chooses the outer loop if the inner loops are short or run briefly. This allows prefetching to happen as early as possible. The system utilizes specific equations based on loop features to ensure that prefetches are only added when they are beneficial. Tests show that APT-GET can significantly improve performance, with speed-ups of up to 1.98× and an average of 1.30× across ten real-world applications. APT-GET is effective, but it has some limitations that typically do not affect its practical use. The main issue comes from the LBR’s small size, which is limited to 32 entries. For example, in nested loops where the inner loop has the delinquent load and runs many times, LBR sampling only captures the inner loop branch. This makes it hard to measure the latency of the outer loop. If the inner loop has more than 32 branches, the LBR only records the critical branch once, which also makes latency measurement difficult. If the loop’s execution time depends on input data, you need to re-profile the application for each input. Still, APT-GET can optimize input-dependent code better than static compile-time methods.

CrossPrefetch~\cite{CrossPrefetch} presents a novel approach to accelerate I/O prefetching, particularly for high-speed storage devices such as NVMe drives. It addresses issues with current OS-level prefetching, such as the rigid interfaces in Linux that do not clearly demonstrate the effectiveness of prefetching, resulting in guesswork and inefficiency. Existing solutions also face performance issues, such as lock contention and inefficient use of system memory. CrossPrefetch employs a cross-layer design, dividing tasks between an OS kernel component (Cross-OS) and a user-level library (Cross-Lib). Cross-OS tracks cached file blocks using a bitmap and shares cache state and prefetch results with Cross-Lib via a new lightweight system call (readahead\_info). If Cross-Lib has this data, they can avoid unnecessary prefetching, keep track of shared file access with detailed indexing (including using range trees), and use aggressive but memory-aware prefetch and eviction policies. This type of collaboration lowers overhead, minimizes cache misses, and alleviates lock contention, while increasing I/O performance (reported to bump from 1.22 to 3.7 times). The existing access pattern predictor uses a simple n-bit counter. It works fine enough in well-tested cases, but it may not efficiently manage more complex or changing I/O patterns. The paper recommends that later studies should investigate more domain-specific predictors. Although the memory used by each file’s bitmap is small, the total memory for many large files, plus the cost of range trees, could add up under tight memory limits. The impact of this on application memory has not been fully tested in extreme cases.

Roland~\cite{Roland} looks at software-based CPU cache prefetching as a key way to reduce memory latency. As CPU speeds and memory access times continue to diverge, especially with the advent of faster DRAM, software prefetching becomes increasingly important. In order to predict complex data structures, such as trees and hash tables, it injects special instructions into the software. The work utilizes micro-benchmarks and a real-world B+ tree example to examine the timing of prefetch instructions and the management of hardware resources, such as the Line Fill Buffer (LFB) and Translation Lookaside Buffer (TLB). The results show that prefetching works best with smaller B+ tree node sizes (such as 256B) compared to larger ones (such as 4kB), and that software prefetching can be used in conjunction with hardware prefetchers to improve data access. In this paper, authors emphasize the need for hardware-aware algorithms and the need for hardware improvement tools such as improved hardware support (i.e., better performance counter details) to make prefetching better. The study notes limitations though, such as that performance differs substantially between CPU architectures owing to the differences of LFB size (10-16 slots on Intel versus 24 on AMD Zen 4) and TLB behavior, which leads to limited portability. It is well established that the analysis is centered around single-threaded applications with limited evaluation of multi-core or multi-socket frameworks. In addition, the use of coroutines for prefetching (as seen in the B+ tree) might add overhead and complexity.

\begin{table*}[h]
\centering
\scriptsize
\begin{tabular}{|p{5cm}|p{4.7cm}|p{2.6cm}|}
\hline
\textbf{Paper Title}  & \textbf{Key Focus}& \textbf{Domain} \\ \hline
 Learning Based Semantic Prefetching& Semantic prediction of DB accesses& Databases~\cite{SeLaP} \\ \hline
Semantic Prefetching using Forecast Slices & Code/runtime memory logic& CPU/memory~\cite{sem1} \\ \hline
Context-Aware Data Prefetching in Mobile Service Environments & Mobile/cloud context& Mobile/Cloud ~\cite{sem2} \\ \hline
Semantic Locality RL Prefetcher & RL/contextual locality& Hardware~\cite{b8} \\ \hline
ContextPrefetcher for Accelerators & Host-guided prefetch& Storage~\cite{b15} \\  \hline
\end{tabular}
\caption{Summary of Semantics and Context Based Prefetching categories}
\label{tab:prefetch_summary}
\end{table*}

\subsection{Semantics Based Prefetching}
Semantic prefetching uses program semantics to predict future accesses to memory while context-based prefetching applies machine and program state to predict future accesses often with reinforcement learning to approximate semantic locality. Semantic-based prefetching can achieve performance increases without much extra hardware cost. Vector architectures (e.g., the RISC-V vector and ARM's Scalable Vector Extension (SVE)) allow programs to specify the total projected data size (called RVL) but hardware limitations often restrict the amount processed per operation (GVL). ReVeLA~\cite{ReVeLA} exploits these RVL-GVL shortfalls to predict the following memory reads, thus providing real-time and accurate prefetch predictions of remaining data (RVL-GVL) prior to the actual utilization. It utilizes a lightweight Stream Tracking Table (STT) as a memory stream monitor, complementing other advanced prefetchers like NextLine(Visual representation in figure:~\ref{fig:nextline}), Best Offset Prefetcher (BOP) and Stream Prefetcher (SPP). ReVeLA boasts a minimum of 436 bytes of hardware overhead and exhibits up to 1.23× better performance when utilized alone and 11.83\% better performing with state-of-the-art prefetchers. One of core innovation is the use of vector code semantics for efficient data cache prefetching.

Conventional prefetchers frequently experience late or unnecessary prefetches due to reliance on reactive, pattern-based heuristics rather than program-level insights. While traditional prefetchers (such as BOP and SPP) achieve high coverage, their limited awareness of vector code semantics hinders their efficiency. In contrast, ReVeLA proactively addresses these shortcomings by leveraging compile-time knowledge of the RVL-GVL gap. However, ReVeLA has notable limitations: (1) it depends on vectorized code and is ineffective for non-vectorized or irregular workloads (e.g., canneal from the multithreaded benchmark suite), as it requires predictable memory access patterns derived from RVL-GVL; (2) it is compatible only with vector-length agnostic (VLA) architectures (e.g., SVE, RISC-V Vector), restricting use with legacy SIMD architectures (e.g., AVX-512); and (3) scalability is limited, as the Stream Tracking Table (STT) supports only 16 entries, which may be insufficient for workloads with many concurrent streams, although the authors note diminishing returns beyond four entries. The underlying principle is that reiterating semantic relationships typically encompasses associated control flows, specific data values, and spatio-temporal patterns. Consequently, monitoring the context of memory accesses may facilitate identification of their semantic location. A context-based neural network (NN) prefetcher can dynamically adapt to arbitrary memory access patterns.

SeLaP~\cite{SeLaP} constitutes a major advancement in data prefetching by transitioning from traditional address-based methods to semantic-aware prefetching, which is more suitable for modern exploratory environments characterized by dynamic and unpredictable query patterns. In contrast to previous systems, which usually cater primarily to navigational (visual) interfaces or structured SQL queries, SeLeP can effectively incorporate many types of workloads so that it serves a wider variety of exploratory data analysis situations. Its architecture is adaptive and learning-based, based on LSTM-based time-series forecasting and dynamic partitioning, that makes the system able to continuously adjust to new workloads rather than relying on static prefetching rules. SeLeP gains considerable performance gains, such as achieving cache hit ratio increases that can rise up to 40\% and an average hit rate of around 95\%, notably greater than the state-of-the-art prefetching techniques. In addition, it decreases I/O wait time by up to 45\%, leading to a better and responsive system for interactive data exploration. Robust assessment on real and benchmarked (SQL-based) workloads with navigational datasets highlights that SeLeP is superior to the baseline approach used, validating both an effective approach and a robust one with different workload environments.

ContextPrefetcher~\cite{b15}, a novel prefetching framework tailored for near-storage accelerators (e.g., computational storage devices with limited DRAM, such as SmartSSDs and Newport CSDs). At the core of ContextPrefetcher is the concept of Cross-Layered Context (CLC) which is a virtual metadata entity bridging host and device. This technique overlooks the hardware/device level cache and exploits host OS level software based caching. They have compared and critisized the Omnicache~\cite{OmniCache} which fails to support prefetching and efficient eviction of the near-storage cache. CLC tracks granular usage patterns of data files, object or a range of object, or range of block aka, block regions (classifying them as active or inactive). ContextPrefetcher uses this to prioritize prefetching active contexts into the device for exclusive cache while evicting inactive ones, ensuring both relevance and freshness of prefetched data. By distinguishing active versus inactive data contexts, the framework supports timely prefetching and eviction—essential for devices with constrained memory budgets.

These data structures have also triggered cache misses in many applications (including parallel trees, graphs, and others) due to pointer chasing, and performance is decimated because memory is accessed unevenly. In the process of solving this problem, Dong~\cite{DTAP} introduces DTAP, a software-hardware co-designed approach that uses type information in strongly typed languages to improve prefetching. The compiler pulls data types and writes metadata to load instructions that allow the prefetcher to recognize the object pointers and also to read directly referenced data, recursively. Prefetch depth is a function of the performance in real-time driven by run-time accuracy and memory bandwidth. Comparing with benchmark specifications, such as SPEC CPU2006 and the C++ STL, DTAP provides 1.37× average speedup compared to traditional prefetching techniques and 5.6× higher coverage because it is type-aware with relatively low accuracy. In this approach the on-chip overhead (1.19 KB) is minimized and the performance is efficient up to multicore systems while aggressive prefetching increases memory bandwidth consumption. By coupling software semantics with hardware efficiency, DTAP can be a useful and low-overhead approach for latency-sensitive applications such as databases and graph analytics, in addition to showing strong performance improvement in memory systems for irregular access patterns. However, DTAP can be faced with problems, including increased memory bandwidth with high levels of aggressive prefetching (up to 90\% higher than some prefetchers) which results in cache contamination and resource contention under constrained bandwidth scenarios. The reliance on compiler-annotated type data restricts it from being applicable in strongly typed languages, and requires refactoring of the programs to fit any legacy code—or those run dynamically. It also faces challenges with deeply nested or dynamically allocated structures, i.e., pointer-to-pointer arrays, in which prefetch depth tuning loses power, yielding low accuracy beyond 5 levels. On the other hand, scalability is limited in applications with extensive data availability or multi-compound types (128 entries per on-chip type table) which may deny critical entries and trigger off-chip table access. In addition, although adaptive depth adjustment addresses some limitations, heuristic-based rules will rarely optimize performance with a variety of workloads, and the choice of simulated environments means actual hardware impacts, such as latency differences are unaudited.

\section{Hybrid Prefetchers}
Purely hardware-based prefetchers are fast and transparent to programmers, but frequently have low semantic knowledge about program behavior, making them less efficient at handling irregular or data-dependent access patterns. Software-based prefetching, on the other hand, employs program knowledge and compiler insights and usually has an instruction overhead and requires careful tuning. To address these complementary limitations, researchers have increasingly explored hybrid prefetching approaches that combine hardware and software mechanisms for responsiveness and programmability. These designs allow software to direct or offer guidance to the hardware prefetcher about promising memory regions, while hardware enables low-latency runtime execution and dynamic adaptation. This collaboration improves prediction accuracy, minimizes unnecessary bandwidth consumption, and increases robustness across diverse workloads~\cite{GRP,PGTP}.

Unlike machine learning (ML) based prefetchers which depend on on-chip learning models alone to identify highly intricate correlations, hybrid strategies provide a unique balance. Hardware ML prefetchers are able to process and dynamically learn irregular access patterns at runtime but usually have high overhead such as storage requirements, energy costs, and latency associated with modeling algorithms. Meanwhile, for hybrid hardware–software prefetchers, some of the intelligence is passed to the compiler or the runtime system, thereby reducing hardware complexity while maintaining semantic knowledge about the program’s behavior. This collaboration allows compilers or software hints to filter out inefficient prefetch candidates and lightweight hardware engines to control precise execution and timing. Accordingly, hybrid prefetching overcomes the scalability and overhead issues of ML-based schemes and enhances adaptability over both regular and irregular workloads~\cite{DEER,HybridPrefetch}.

A recent study shows that combining software guidance with hardware execution increases prefetching efficiency significantly. Prophet~\cite{PGTP} introduces a profile-guided temporal prefetcher that uses lightweight performance counter profiling to inject compiler-generated hints into binaries and allow hardware to make more accurate insertion, replacement, and resizing decisions. This mixed model not only surpasses state-of-the-art hardware prefetchers such as Triangel~\cite{ref10} by more than 14\%, but also demonstrates how software-provided semantic insight can reduce metadata overhead while maintaining the responsiveness of hardware prefetching. DEER~\cite{DEER} leverages profile analysis in software to extract metadata and guide hardware to prefetch instruction cache lines well in advance, which is especially useful for mobile workloads with large call stacks. It achieves significant improvement while maintaining low on-chip overhead.

\subsection{Hardware-assisted software prefetching with runtime feedback}

This approach is based on hardware mechanisms that recognize memory access patterns and trigger prefetches via consistent feedback. HoPP (Hardware-Software Co-designed Page Prefetching)~\cite{hopp} is a framework for prefetching pages that is designed for disaggregated memory architectures. It alleviates the semantic gap between applications and the operating system by utilizing a system-level memory access trace collection mechanism at the memory controller to better predict hot pages in the operating system. With an adaptive three-tier prefetching mechanism, HoPP finds several access patterns in parallel, for example simple sequential streams, ladder, and ripple patterns, prefetching pages asynchronously to hide remote memory latency. The hardware support for trace capture, including a hot page detection module and a reverse page table cache at the memory controller, are part of the design. On the software side, HoPP’s prefetch engine dynamically manages what the prefetch algorithms look like, and injects page table entries early on to avoid page-fault overheads. While the advantages of the approach can be offset by increases in complexity: specialized hardware modules lead to the system being more complex (and consuming more energy), and effectiveness lies in the accurate identification of stream patterns. Additionally, HoPP's large-scale deployment requires full memory trace collection infrastructure, leading to significant overhead and scalability issues.

Record \& Replay (RnR)~\cite{RnR} is a hardware prefetcher guided by software hints, designed to exploit repetitive memory access patterns in irregular applications. Programmers annotate code to specify target data structures and iteration boundaries where irregular access sequences are likely to repeat. During a designated record phase, such as the first iteration of a loop, RnR captures the sequence of cache-missing addresses. In subsequent iterations, it replays the recorded miss sequence as prefetch requests, anticipating recurrence of the same accesses. This software-marked record/replay approach enables RnR to manage long, complex access patterns that conventional spatial or temporal hardware prefetchers cannot predict effectively. RnR leverages programmer insight about temporal (iteration) and spatial (data structure) regions to prefetch precisely what will be needed in the next iteration. By prefetching an exact previously observed address stream, RnR achieves exceptionally high prefetch accuracy and miss coverage. In evaluations of irregular workloads, RnR attained over 95\% prefetching accuracy and covered more than 90\% of cache misses. This performance significantly surpasses traditional spatial-correlation or temporal-correlation prefetchers, which often retrieve only a fraction of the required data in irregular patterns. General-purpose hardware prefetchers, such as Bingo~\cite{b7} and SteMS~\cite{STMS}, typically experience accuracy drops below 50\% on pointer-heavy or graph workloads. In contrast, RnR replays the exact miss addresses in the correct order and timing, resulting in nearly every prefetched block being used (yielding approximately 97\% useful prefetches). This leads to high miss coverage and minimal incorrect prefetches, producing substantial speedups (e.g., 2.16× in graph analytics, 2.91× in an iterative sparse solver).

A lightweight ML-based runtime prefetcher~\cite{ref11} selection technique employs a supervised learning model with phase-based classification of hardware performance counter metrics to dynamically enable or disable components of composite prefetchers. Implemented as a compact decision tree (requiring only 42 bytes of storage), this approach was validated on a 160-core Ampere server platform and achieved speedups of up to 25\% by adjusting prefetchers to match workload phases. Its effectiveness notwithstanding, the approach does have a few weaknesses. It is sensitive to the representativeness of its training data and changes in workload phase, fails to adapt per-core, and may degrade performance for workloads with unseen or highly dynamic behaviors. Cache pollution or inter-core interference side effects are not adequately considered by the evaluation. The method is dependent on hardware PMU availability and periodic decision tree inference, which may restrict responsiveness. Further, the offline training and trace collection overhead is considerable, presenting practical deployment challenges, and the fixed decision tree model structure is not suitable for fine control or generalization in real production applications.

\subsection{Software-guided hardware prefetching}
Software-guided hardware prefetching uses software hints to influence hardware prefetching decisions and optimize prefetch targets. Ainsworth~\cite{ProgP} presents an event-triggered programmable prefetcher tailored for irregular memory access patterns, that are commonly found in graph, database, and high-performance computing workloads. While traditional prefetchers perform poorly with complex and unpredictable memory traversals, this type utilizes programmable logic units (PPUs) that are activated by compiler-generated events triggered by memory accesses. These PPUs can execute small computation kernels derived from annotated source code, enabling advanced and timely prefetching without stalls and achieving up to a 3.0× performance improvement.

While this has its own advantages, it has some critical drawbacks. Its effectiveness also relies on the quality of the compiler-generated annotations and the efficiency of PPU event scheduling and may not cope with dynamic changes and workload-specific intricacies effectively. Also, this method involves hardware-specific changes, requiring additional PPUs and event-triggering infrastructure; this introduces concern for increased chip complexity, power consumption, and silicon area overhead (approximately 3\%). Static annotation may also limit adaptability in highly dynamic or unforeseen runtime scenarios and hinder broader adoption on general-purpose processors.

GRP~\cite{GRP} is a hybrid hardware-software prefetching approach that leverages compiler-generated memory access hints with a hardware prefetch engine to increase memory performance. It extends Scheduled Region Prefetching (SRP), which utilizes spatial locality, allowing the compiler to annotate load instructions using semantic cues, including spatial locality, pointer dereferences, recursive structure traversals, and indirect array accesses. These annotations direct the hardware-triggered prefetches upon L2 cache misses. GRP solves major constraints of only hardware prefetchers, including the poor accuracy due to blind speculation, and software-only schemes, like low scheduling depth and execution overhead. Prefetches are dynamically scheduled, but they are context-sensitive—restricted to a specific task at a given time through static analysis, thus avoiding the use of deep runtime learning tables or speculative execution hardware.

GRP has a number of limitations, such as its effectiveness being limited by static analysis on the behavior of memory access that fails to generalize when working on dynamic or input-dependent applications. The GRP supports pointer-based and recursive accesses, but benefits are limited for non-affine access patterns, such as hash tables or linked data structures with inconsistent traversal depths. As part of the overall framework, the prefetch engine will need to read/interpret hints, maintain prefetch queues, and also manage recursive pointer chaining using hardware recursion counters that is an increase in complexity within the microarchitecture. Despite GRP optimizations some benchmarks, such as mcf, art, etc., are performance limited by bandwidth saturation or lack of prefetchable regularity. Encoding hints through unused instruction bits or novel opcodes often makes ISA design more difficult or causes interferences with pre-existing software toolchains. In conclusion, GRP provides a trade-off between software accuracy and hardware aggressiveness but its dependence on compiler-hinted regularity and the hardware integration need restricts its scalability for highly irregular or dynamically adaptive applications.

This multi-strided access pattern~\cite{multistridedaccesspatternsboost} approach to hardware prefetching is proposed to transform memory-bound loops in order to make multiple successive memory strides to be run concurrently in each iteration by either loop unrolling or interchange, and the method of doing so activates multiple hardware prefetchers. With up to 12.55× memory throughput faster than the state of the art libraries Polly, MKL, and OpenBLAS~\cite{OpenBLAS}, this multi-strided approach increases throughput. But it requires regular and dense memory access patterns and could conflict against the cache set if too many strides map to the same cache set. The method also raises register pressure and thus making stride configurations requiring more registers than are available infeasible, and this is not suitable to irregular kernels that require a gather/scatter memory access. At a practical level, multi-striding now needs some low-level code generation and tuning of unroll factors (via hand-tuned AVX2 assembly), as there has been no compiler support. This causes portability and tuning issues that become crucial for multi-core or heterogeneous systems that expand beyond the tested single-core scenarios.

\section{Moving to Complex Pattern Based Prefetching}
The primary motivation for using complex pattern-based prefetching is the inefficiency of traditional methods in managing the spatial distribution of block access requests. According to Piatov~\cite{ref17}, existing prefetching techniques often ignore the interleaved access patterns generated by multithreading and higher memory hierarchy layers, leading to suboptimal performance. To tackle these challenges, the DeepPrefetcher model was proposed, using deep neural networks to adaptively learn and predict access patterns. Overall, the model realized significant improvements, surpassing traditional prefetching methods by 21.5\% in accuracy and 17.2

In solid-state drives (SSDs), Li~\cite{ref16} proposed a pattern-based prefetching scheme operating at the flash translation layer. By detecting common block access patterns while implementing adaptive cache management, this method significantly reduced read latency and kept write latency consistent. This improvement showcases the opportunity to benefit from pattern recognition and the use of cache management for improving performance of storage systems, helping to transition to more complex prefetching methods.

The need for advanced prefetching solutions is further emphasized by other challenges faced by virtualized environments. Conventional memory prefetching methods like Leap, according to Wang~\cite{ref14}, often prove to be inadequate in virtual machine settings due to repeated memory resource usage. Their work justifies the use of spatial-temporal pattern-based prefetching, which satisfies the increasing requirement of optimized memory handling in cloud applications. Such method enhances resource utilization and preserves service quality, further substantiating the case for complex pattern-based strategies.

Secondly, the use of deep learning-based methods in cache management has generated promising results. Wang~\cite{ref15} proposes a CNN-LSTM model that integrates convolutional neural networks' ability of spatial feature extraction and temporal modeling ability of long short-term memory networks. This hybrid model has shown great benefits for predicting cache demand in comparison to classic algorithms like LRU and LFU. The ability to dynamically optimize cache allocation based on predicted needs is crucial for enhancing system response times and resource efficiency.

The increasing complexity of memory access patterns not only reveals the limitations of conventional heuristics, but also the challenges of lightweight machine learning models like simple perceptrons or decision trees. Although these models can deal with only linear or shallow correlations, they suffer from long-range dependencies, nested structures, and multi-level irregularities usually encountered in graph analytics, sparse matrices, and dynamic pointer-based workloads. In order to capture these higher-order relationships, researchers have naturally turned to deep learning models which are well-suited for sequence modeling and feature extraction. Delta-LSTM and CNN-LSTM methods incorporate techniques using recurrent and convolutional architectures to learn temporal evolution and local spatial features in address streams, which enables prefetchers to predict nontrivial patterns that shallow models cannot effectively capture. This is an indication of the general trend in computer architecture towards using deep learning not only for application domains like vision and NLP, but also as a powerful tool for uncovering hidden structure in hardware access traces.

A key limitation of cache prefetching is the complexity used in pattern-based cache retrieval, which will be addressed. To enhance the performance of data retrieval and processing we adopt specialized machine learning algorithms, control the cache adaptively and interpret dynamically evolving data access patterns. With increasing data volumes and increasing computing resource requirements, advanced prefetching algorithms will be crucial to improve performance in modern storage systems. We have reviewed existing literature for a strong basis for continued research and innovation in this important field.

\subsection{Predicting non-uniform access pattern}

Predicting non-uniform access patterns remains a significant challenge in prefetching research because these patterns often arise from irregular control flow, data-dependent accesses, or dynamic program behavior that resists static analysis. In contrast to uniform access patterns, such as regular strides or simple temporal locality, non-uniform patterns frequently appear in applications including graph analytics, sparse matrix computations, and pointer-based traversals. 

In these scenarios, memory accesses are governed by complex data relationships or runtime input. Recent studies have investigated machine learning and history-based approaches to model these irregular behaviors and enhance prefetching accuracy. For instance, Wenisch~\cite{ref21} introduced correlation-based prefetching to capture non-linear dependencies in memory accesses. Jain and Lin~\cite{ref22} proposed neural network-based prefetchers that learn from historical memory reference streams to address complex and dynamic patterns. The following section reviews representative techniques for non-uniform access patterns, highlighting their respective contributions and limitations. Classical data prefetchers usually find sequential or frequently spaced (strided) memory access patterns. More recent prefetchers use delta correlation-based methods that recognize and predict recurrent patterns in access intervals rather than constant ones. When applications access different elements in large data structures, non-uniform deltas within a page are formed. Delta correlation is effective only when access patterns are observed consistently during program execution. If the working set is more irregular, advanced algorithms would need to be used for accurate prediction. A common approach is to understand the intrinsic knowledge of the software to locate relevant elements for prefetching. Prefetchers may be implemented at various levels in the cache hierarchy. In high-throughput systems, prefetcher position drastically affects overall performance~\cite{VLDP}.

Shevgoor~\cite{VLDP} uses a history table to investigate complex patterns in the access stream. One such method introduced by the author is known as Variable Length Delta Prefetcher (VLDP) which mitigates prefetching delays. In most cases, this prefetcher is suitable for the last-level caches which predicts the page stream from memory. A small structure, Delta History Buffer (DHB) simply keeps their own local histories for every physical page in the current workload. If the prefetching opportunity is discovered and a corresponding delta history is detected, this history is queried in the Delta Prediction Table (DPT) — a sequence of key-value pairs that connects all previous delta patterns to the next expected access delta within a page. VLDP prioritizes predictions from DPT tables tracking longer histories so as to mitigate aliasing and ensure accuracy and cascades to the downstream prediction in a similar way - which is basically using multiple DPT tables having differing history length of history.

Temporal Memory Streaming (STMS)~\cite{STMS} is designed to efficiently manage off-chip metadata storage. STMS uses Hash-Based Lookup, which employs a hardware-managed main-memory hash table to minimize latency in indexing miss-address sequences. The probabilistic Update reduces memory bandwidth by randomly sampling metadata updates, trading minimal accuracy for substantial bandwidth savings. Amortized Lookups structure metadata so each off-chip lookup predicts multiple subsequent misses, distributing the lookup cost. By storing metadata in main memory, STMS significantly reduces on-chip storage demands while retaining about 90\% of the performance achievable with idealized on-chip metadata storage. This offers a practical balance of performance, storage efficiency, and memory bandwidth usage.

The Irregular Stream Buffer (ISB)~\cite{Temporal4} is a prefetcher that linearizes the physical addresses associated with patterned irregular access in memory to allow faster access. This approach allows accurate prediction of repeatedly correlated address sequences, performance enhancement on irregular SPEC benchmarks and the prefetch accuracy remains at 93.7\%. The principle of the ISB being that address sequences tend to repeat as the main one and hence transforming the prediction of temporal patterns from complicated table lookups to sequential predictions in terms of the structural indirection. It uses two Address Mapping Caches (AMCs): the Physical-to-Structural (PS-AMC) and the Structural-to-Physical (SP-AMC) ones, to preserve bidirectional address mappings. Upon meeting an address trigger, this system converts it to the structural space so that predictor finds the next address of the sequence, and then converts it back into the physical space for prefetching. The Training Unit records temporal correlations from the program counter (i.e., load instruction), and then distributes instruction data to create coverage and accuracy. On-chip metadata is compressed to 13-bit and cached only for TLB-resident pages, resulting in memory traffic overhead of 8.4\% versus 35\% of alternatives. It reduces memory overhead and maximizes performance by leveraging temporal patterns that come from recent accesses, as opposed to storing everything from the past.

The Managed Irregular Stream Buffer (MISB)~\cite{MISB} is an advanced temporal prefetcher designed to exploit temporal locality by efficiently managing metadata for irregular data access patterns. MISB achieves a 22.7× speedup on irregular workloads and reduces metadata traffic overhead by 5.9× compared to previous methods. The primary challenge addressed by MISB is the efficient management of the substantial metadata required to track temporal correlations. The central innovation lies in the recognition that metadata can be cached effectively if organized at an appropriate granularity. MISB utilizes two on-chip metadata caches: the PS Cache (Physical-to-Structural), which maps physical addresses to structural addresses, and the SP Cache (Structural-to-Physical), which provides the inverse mapping for prefetch generation. In contrast to prior approaches such as ISB, which cache metadata at the coarse granularity of entire TLB-page mappings (resulting in 90\% unused data), MISB caches individual 8-byte mappings, achieving a 66\% hit rate compared to ISB's 30\%. This fine granularity supports temporal filtering by retaining only temporally relevant mappings. Since metadata also exhibits temporal locality, MISB incorporates a metadata prefetcher that applies next-line prefetching in the SP cache. Analysis reveals that 65\% of PS cache misses correspond to requests for previously unseen physical addresses, which results in wasted bandwidth. To address this, a Bloom filter is employed to track which physical addresses have associated mappings. ISB's approach of TLB-synchronized metadata caching is limited because it fetches an entire page's metadata, of which only approximately 10\% is utilized, making it unsuitable for 2MB pages due to the excessive on-chip cache requirements (200-400KB). This method also lacks metadata prefetching and can introduce latency stalls. Similarly, in STMS and Domino, the significant traffic overhead from large GHB-based metadata structures leads to poor temporal locality in history buffer access patterns. In contrast, MISB organizes metadata for cacheability, as structural-to-physical mappings demonstrate strong spatial and temporal locality. This approach proactively hides latency, while the Bloom filter eliminates unnecessary requests.

\subsection{Using Machine Learning for prediction }

Machine learning in general to predict multi-layered non-uniform patterns of access pattern has emerged to be a promising area of prefetching research and will provide an answer to traditional hardware and heuristic-based processes. Compared to a simple stride- or history-centric prefetching approach, ML-based methods can be used to discover complicated correlations, long-range dependencies, and context-dependent processes in the memory access stream. This can be advantageous particularly for workloads with non-regular and data-driven access patterns such as graph analysis, sparse computing and pointer-dominated workloads. Techniques such as recurrent neural networks (RNNs), perceptron-based models, and reinforcement learning networks have been used in programming to accommodate dynamic program phases and access behaviors. Srinivasan et al. (2019) presented a perceptron based prefetcher that enhances the prediction capability of a complex pattern ~\cite{ref23} while Hashemi et al. (2018) introduced learning-based techniques to model erratic memory behavior in multiple applications~\cite{Delta-LSTM}. More recently, Bhattacharjee et al. (2022) studied Graph Neural Networks (GNNs) to learn the memory access patterns in graph workloads~\cite{ref25}. The main advantage of ML-based prefetching is its generalization of different access patterns without any explicit requirements from programmers or compiler, and therefore leads to more intelligent, adaptive, and workload independent prefetching. Representative ML-based prefetching algorithms are covered below and their contribution towards managing the complexity of non-uniform access patterns is presented.

Delta-LSTM~\cite{Delta-LSTM} addresses the challenge of memory prefetching in modern computing systems by targeting the complexity of irregular memory access patterns. This approach utilizes recurrent neural networks (RNNs) with Long Short-Term Memory (LSTM) units, referred to as DELTA-LSTM, to prefetch data by learning and predicting memory access patterns. The method reframes the memory prefetching problem as a sequence classification task, discretizing memory address deltas to manage the vast and sparse output space. By clustering addresses to reduce complexity, DELTA-LSTM achieves higher precision and recall than traditional stride and correlation-based hardware prefetchers, effectively capturing complex and irregular patterns.

The Compressed-LSTM~\cite{CompressedLSTM} approach, which accurately predicts using traditional machine learning models. The authors introduce a novel compression technique based on output encoding. For predicting one of n possible memory locations, their approach reduces the model's parameters from a dense layer of size n to a binary representation of size \textit{log}n. This results in a significant compression factor of approximately O(n/log n), achieving approximately 100× compression without a substantial loss of accuracy.  The highly compressed LSTM models are viable and advantageous for predicting memory accesses. 

RAOP~\cite{RAOP} leverages past memory access patterns using a Recurrent Neural Network (RNN) with Long Short-Term Memory (LSTM). Rather than analyzing only memory addresses, RAOP monitors the differences (offsets) between consecutive memory accesses to identify latent patterns. The LSTM model then predicts memory addresses in the future with greater accuracy than traditional prefetchers. RAOP interacts directly with the cache memory system of computer architecture as a part of its hardware-level implementation. In modern processor designs, prefetching reduces CPU idle time waiting for data. RAOP significantly improves prefetching accuracy by 3.22 times, memory coverage by 4.2 times, and the overall system efficiency is enhanced by 15.6\%, thereby achieving a considerable increase in computational efficiency. The results indicate that AI can enhance software and hardware performance, leading to the advancement of smart computing systems. Despite these benefits, RAOP has a few drawbacks where it requires insignificant computation power and is complex in connection with agentic AI. The system needs to be trained for individual workloads and it could require retraining if workload characteristics shift with new workloads. Prediction delays can reduce the model's effectiveness, and its memory usage may pose challenges for devices with limited resources. RAOP is not efficient enough particularly when the resource allocation is based on random memory access patterns as it requires too much extra overhead.

Voyager~\cite{voyager} is a neural-based prefetcher that is comparable to ISB.  With 19.4\% higher coverage in learning temporal correlations relative to ISB, it has the ability to capture both temporal and spatial memory access patterns. Even without the use of delta information (Voyager w/o delta). Voyager more effectively predicts spatial (56.8\% vs. 45.2\%) and non-spatial (22.2\% vs. 13.1\%) access patterns than ISB. Analysis of uncovered patterns reveals that Voyager reduces most categories of uncovered accesses, except for compulsory misses. Notably, Voyager has the ability to mitigate compulsory misses by incorporating common delta information into its predictive model. The author has given a scope of its adaptability and the potential for even greater improvements.

Spiking neural networks in Pathfinder~\cite{b10} enable real-time data prefetching by effectively learning and anticipating access patterns, thereby addressing the limitations of conventional neural networks in prefetching tasks. Each data access pattern varies, requiring a distinct set of rules for each major pattern type. The study proposes that the neural prefetcher offers a concise formulation that encapsulates multiple rules as patterns recognizable by neurons with adjustable weights. The researchers employed spike-timing-dependent plasticity, a biological mechanism that modulates synaptic strength between neurons in the brain, as the basis for their hypothesis.

Bhati's~\cite{PPF}(Perceptron-based Prefetch Filtering) aims to increase the coverage of an underlying prefetcher without negatively impacting accuracy. PPF is implemented as an enhancement to Signature Path Prefetcher(SPP). Look-ahead prefetchers, such as SPP, provide a mechanism to speculate an arbitrary number of references ahead of the initial triggering access. A throttle confidence threshold is then used to ensure that the lookahead stops when confidence falls too low to ensure that prefetches are accurate. However, as the lookahead depth increases, so do useful prefetches and hence coverage. So this challenge of increasing coverage without affecting accuracy is accomplished by PPF. PPF allows SPP to speculate deeply to achieve high coverage while filtering out the inaccurate predictions that this deep speculation implies. In this design, PPF replaces SPP’s confidence-based throttling mechanism. Since PPF is more effective in rejecting inaccurate prefetches, the SPP’s design can be re-tuned to achieve maximum coverage, hence providing an increase in both accuracy and coverage and a notable increase in performance. On a memory intensive subset of the SPEC CPU 2017, in a single core configuration, PPF increases performance by 3.78\% compared to SPP and by 11.4\% for a 4-core configuration. The Perceptron Learning Model (A machine learning model) implemented by the PPF makes use of a Perceptron Filter for filtering out the inaccurate prefetch. This filter needs to be trained to avoid mispredictions. This process requires time and added computational complexity, which may not be ideal for systems with limited capacity. For implementing this on a larger scale, larger or multiple perceptrons might be required, which may require additional memory due to which storage overhead can be another issue. Additionally, although the filter is designed to adapt to memory access patterns, it might struggle with unpredictable or highly dynamic workloads, leading to mispredictions.

Bingo~\cite{b7} is a prefetcher that generates requests based on spatially linked data access patterns. It is designed to optimize data access efficiency in applications with consistent memory layouts.Conventional spatial prefetchers are limited by their bias toward either short, frequent but less accurate events or long events that are precise but infrequent, resulting in missed prefetch opportunities. Bingo solves these limitations by utilizing a dual-event strategy that correlates spatial patterns with both short and long events. This dynamic approach improves prefetch accuracy and minimizes cache misses through the selection of the optimal access pattern, which is therefore most relevant. With an efficiently scalable structure, Bingo takes advantage of a single history table to store event associations, which adds to overall performance. Analysis in a variety of big-data applications shows that system performance is 60\% better than prefetching-free systems, and 11\% better than top-of-the-line spatial prefetchers. By combining short and long event correlations, we achieve intelligence-driven and timely data prefetching. However, the Bingo Spatial Data Prefetcher also has its own limitations. It is well intended but only works if event associations are accurate, where incorrect correlation can cause bad prefetching and unwanted memory accesses, as well as cache pollution. While a single history table is efficient and less time consuming for multiple use cases, scalability becomes more difficult to cope with the variety of data patterns as the access options become more varied. After performing the average scaling of more than 16K entries, we can see diminishing returns, or a plateau in miss coverage. In addition, it is important to keep a healthy balance between short events and long events so that an imbalance lowers predictive accuracy and limits adaptability with applications accessing different patterns.

It employs the TAGE-prefetcher~\cite{b14}, an indirect branch predictor at its core. Context-Based Prefetcher~\cite{b8} employs program semantics to improve performance. A reinforcement learning-based approach using program characteristics for the prefetcher has been presented. 

Pythia\cite{pythia}, which formulates hardware prefetching as a reinforcement learning problem. Reinforcement learning is a machine learning paradigm that studies how an autonomous agent can learn to take optimal actions that maximize a reward function by interacting with a stochastic environment. The Paper formulates Pythia as an RL-agent that autonomously learns to prefetch by interacting with the processor and the memory subsystem. For every new demand request, Pythia extracts a set of program features. It uses the set of features as state information to take a prefetch action based on its prior experience. For every prefetch action (including not to prefetch), Pythia receives a numerical reward which evaluates the accuracy and timeliness of the prefetch action given various system-level feedback information. Using the RL framework, Pythia can holistically learn to prefetch using both multiple program features and system-level feedback information inherent to its design. Pythia can also be easily customized in silicon via simple configuration registers to exploit different types of program features and/or change the objective of the prefetcher. This gives Pythia the unique benefit of providing even higher performance improvements for a wide variety of workloads and changing system configurations, without any changes to the underlying hardware. Pythia outperforms two state-of-the-art prefetchers (MLOP and Bingo) by 3.4\% and 3.8\% in single-core, 7.7\% and 9.6\% in twelve-core, and 16.9\% and 20.2\% in bandwidth-constrained core configurations, while incurring only 1.03\% area overhead over a desktop-class processor and no software changes in workloads. The limitations mentioned are implied limitations and not explicitly mentioned in the research paper. While the paper claims that Pythia is general enough to incorporate various system level feedback information, it just demonstrates one major system level feedback information for prefetching: Memory Bandwidth Usage. Pythia uses reinforcement learning, meaning it learns as the program executes. While this reduces the need for offline training, it can introduce a learning phase where the prefetcher is not yet effective. During this initial phase, performance might be lower compared to traditional prefetchers with fixed algorithms.

\subsubsection{\textbf{Taxonomy of ML Approaches for Prefetching}}
Machine learning techniques applied to memory access prediction can be organized along two principal axes: the learning paradigm (supervised, reinforcement, or unsupervised/semi-supervised) and the training mode (online versus offline). This dual-axis classification captures both what the model learns and when it learns, which together determine the model's suitability for deployment at different levels of the cache hierarchy.

Supervised learning methods train on labeled pairs of memory access histories and their corresponding next addresses (or deltas). Recurrent architectures such as Long Short-Term Memory (LSTM) networks are the dominant choice in this category, because the prefetching problem is inherently sequential: given a history of recent accesses, predict the next one. Delta-LSTM~\cite{Delta-LSTM} reformulates memory prefetching as a sequence classification task over discretized address deltas, while Compressed-LSTM~\cite{CompressedLSTM} applies output encoding to achieve approximately 100× compression with minimal accuracy loss. RAOP~\cite{RAOP} extends this line by monitoring offsets rather than raw addresses. Beyond recurrent models, Bhattacharjee et al.~\cite{ref25} employ Graph Neural Networks (GNNs) to capture topology-aware dependencies in graph-dominated workloads, and Wang et al.~\cite{ref15} fuse convolutional and recurrent layers (CNN-LSTM) to extract both spatial and temporal features in cache demand streams. Voyager~\cite{voyager} introduces a hierarchical neural architecture that captures correlations at multiple granularities, achieving 19.4\% higher coverage relative to the Irregular Stream Buffer. While supervised methods offer high model capacity, they require offline training on representative traces and are sensitive to distribution shifts when workload characteristics change.

At the lighter end of supervised learning, perceptron-based models deserve separate attention due to their fundamentally different deployment profile. The Perceptron-based Prefetch Filter (PPF)\cite{PPF} uses a simple linear classifier not to generate prefetch addresses directly, but to filter out inaccurate candidates produced by an underlying look-ahead prefetcher (SPP). With only ~13 KB of storage, PPF operates online with per-access weight updates, making it one of the most hardware-amenable ML-based approaches. Similarly, the lightweight decision tree model of Alcorta et al. ~\cite{ref11}, requiring just 42 bytes, performs runtime prefetcher selection based on hardware performance counter metrics. These models trade model capacity for immediate deployability.
Reinforcement learning (RL) formulates prefetching as a sequential decision problem. An RL agent observes program state features, takes a prefetch action (including the option of not prefetching), and receives a reward signal based on the accuracy and timeliness of its decision. Pythia\cite{pythia} is the most prominent example, framing hardware prefetching as an RL problem where the agent learns from multiple program features and system-level feedback (specifically memory bandwidth usage). Pythia achieves significant gains over state-of-the-art prefetchers (e.g., 16.9–20.2\% improvement over MLOP and Bingo in bandwidth-constrained configurations) with only 1.03\% chip area overhead. The context-based prefetcher of Peled et al.~\cite{sem1} takes a related approach, using reinforcement learning to approximate semantic locality. The key advantage of RL is its ability to optimize for a holistic objective (e.g., balancing accuracy, timeliness, and bandwidth) rather than a single loss function. However, RL agents require a learning warm-up phase during which performance may be suboptimal, and the demonstrated system feedback has so far been limited to bandwidth.

Unsupervised and semi-supervised approaches discover latent structure in access streams without requiring explicit next-address labels. The clustering stage in Delta-LSTM~\cite{Delta-LSTM} groups addresses to reduce the output space dimensionality before supervised prediction. SeLeP ~\cite{SeLaP}  uses dynamic partitioning combined with LSTM-based time-series forecasting for semantic-aware prefetching in exploratory database workloads, achieving cache hit ratios up to 95\%. Pathfinder ~\cite{b10}, based on spiking neural networks (SNNs) with spike-timing-dependent plasticity (STDP), represents an emerging bio-inspired paradigm where learning occurs through event-driven weight updates—inherently unsupervised and energy-efficient. While these methods reduce the dependence on labeled training data, their accuracy can be sensitive to the quality of clustering or partitioning decisions.

\begin{table*}[htbp]
\centering
\scriptsize
\renewcommand{\arraystretch}{1.3}

\begin{tabular}{|p{2cm}|p{1cm}|p{2cm}|p{3.5cm}|p{3.5cm}|}
\hline
\rowcolor{blue!60}
\color{white}\textbf{ML Paradigm} & 
\color{white}\textbf{Learning Mode} & 
\color{white}\textbf{Representative Techniques} & 
\color{white}\textbf{Strengths} & 
\color{white}\textbf{Limitations} \\
\hline

\rowcolor{blue!10}\multicolumn{5}{|l|}{\textit{\textbf{Supervised Learning-Learns input-output mappings from labeled \(address, next-address\) pairs}}} \\
\hline

RNN / LSTM 
& Offline (pre-trained on traces) 
& Delta-LSTM ~\cite{Delta-LSTM} , Compressed-LSTM~\cite{CompressedLSTM}, RAOP ~\cite{RAOP}  
& Captures long-range temporal dependencies; handles irregular delta sequences; compressible for deployment 
& High training cost; large model size without compression; offline retraining needed for workload shifts \\
\hline

Perceptron / Linear 
& Online (updated at runtime) 
& PPF~\cite{PPF} , Context-Based ~\cite{sem1}  
& Extremely low inference latency; minimal storage ($\sim$42 B--13 KB); real-time adaptation; suitable for on-chip deployment 
& Limited to linear/shallow correlations; cannot capture higher-order dependencies or nested structures \\
\hline

Graph Neural Network 
& Offline 
& Bhattacharjee et al. ~\cite{ref25} 
& Naturally models graph-structured access patterns; captures topology-aware dependencies in graph analytics 
& Requires graph construction overhead; high training complexity; limited to graph-dominated workloads \\
\hline

CNN-LSTM Hybrid 
& Offline 
& Wang et al. ~\cite{ref15} 
& Combines spatial feature extraction (CNN) with temporal modeling (LSTM); strong for cache demand prediction 
& Dual-network overhead; not validated on real hardware; complex pipeline may add latency \\
\hline

Hierarchical Neural 
& Offline 
& Voyager ~\cite{voyager}  
& Captures both temporal and spatial correlations via multi-level architecture; 19.4\% higher coverage over ISB 
& Multi-level model complexity; requires substantial training data; not yet demonstrated at scale on-chip \\
\hline

Spiking Neural Network 
& Online (STDP) 
& Pathfinder ~\cite{b10} 
& Biologically inspired; real-time learning via spike-timing-dependent plasticity; event-driven (energy-efficient) 
& Emerging paradigm with limited tool support; accuracy vs.\ conventional NNs not fully benchmarked \\
\hline

\rowcolor{blue!10}\multicolumn{5}{|l|}{\textit{\textbf{Reinforcement Learning-Agent learns optimal prefetch actions via reward signals from the memory subsystem}}} \\
\hline

Tabular / Q-Learning RL 
& Online 
& Peled et al. ~\cite{sem1}  
& Approximates semantic locality; adapts to context-dependent patterns without labels 
& State-space explosion for large address spaces; slow convergence on complex workloads \\
\hline

Deep RL 
& Online 
& Pythia ~\cite{pythia}  
& Holistic learning from multiple program features and system-level feedback; customizable via configuration registers; 1.03\% area overhead 
& Learning warm-up phase with suboptimal performance; demonstrated only with bandwidth as system feedback \\
\hline

\rowcolor{blue!10}\multicolumn{5}{|l|}{\textit{\textbf{Unsupervised/Semi-Supervised-Discovers latent structure in access streams without explicit labels}}} \\
\hline

Clustering + LSTM 
& Offline / Semi-supervised 
& Delta-LSTM ~\cite{Delta-LSTM}  (clustering stage), SeLeP ~\cite{SeLaP}  
& Reduces output space via address clustering; enables semantic-aware prediction for exploratory workloads 
& Clustering quality directly affects prediction; may lose fine-grained address resolution \\
\hline

Decision Tree (Phase Classification) 
& Online (lightweight) 
& Alcorta et al.~\cite{ref11}
& 42-byte model; phase-based prefetcher selection at runtime; validated on 160-core server 
& Sensitive to training data representativeness; no per-core adaptation; fixed tree structure limits generalization \\
\hline

\end{tabular}

\caption{Machine Learning Paradigms for Prefetching}
\label{tab:ml_prefetch}
\end{table*}

\subsubsection{Comparative Analysis of ML-Based Prefetchers}
While the taxonomy in Section 6.2.1 classifies techniques by their algorithmic foundations, practical evaluation requires a multi-dimensional comparison that accounts for quantitative performance, resource costs, and deployment constraints. Table 6 provides this comparison across all ML-based prefetchers discussed in this survey. The table is organized chronologically to also reveal the evolution of the field from early offline LSTM models toward lighter, online-capable designs.

Several patterns emerge from this comparison. First, there is a clear accuracy–overhead trade-off: LSTM-based methods (Delta-LSTM, RAOP, Voyager) achieve the highest raw accuracy and coverage improvements but carry storage and compute costs that preclude direct on-chip deployment without compression. In contrast, perceptron-based (PPF) and RL-based (Pythia) models sacrifice some model expressiveness for dramatically lower overhead, making them the only ML prefetchers demonstrated with practical on-chip area costs. Second, the training mode critically shapes deployment feasibility: offline models require trace infrastructure and retraining pipelines, while online models (PPF, Pythia, Pathfinder) adapt in real time but may underperform during initial learning phases. Third, no single technique dominates across all dimensions, which motivates the hierarchical or ensemble strategies we discuss in Section 6.2.3.

\begin{table*}[htbp]
\centering
\scriptsize
\renewcommand{\arraystretch}{1.3}

\begin{tabular}{|p{1cm}|p{1cm}|p{1.5cm}|p{1.5cm}|p{1.5cm}|p{2cm}|p{3cm}|}
\hline
\rowcolor{blue!60}
\color{white}\textbf{Prefetcher} &
\color{white}\textbf{ML Model} &
\color{white}\textbf{Storage Overhead} &
\color{white}\textbf{Accuracy / Coverage Gain} &
\color{white}\textbf{IPC / Speedup} &
\color{white}\textbf{HW Feasibility} &
\color{white}\textbf{Key Limitation} \\
\hline

Delta-LSTM ~\cite{Delta-LSTM}  & LSTM (RNN)  
& $\sim$4 MB (full); $\sim$40 KB (compressed) 
& Higher precision \& recall vs.\ stride 
& Moderate 
& Low (GPU training; large uncompressed model) 
& Vast output space; offline retraining for new workloads \\
\hline

Compressed-LSTM ~\cite{CompressedLSTM} & LSTM + output encoding  
& $\sim$100$\times$ compressed ($\sim$40 KB) 
& Comparable to full LSTM 
& Moderate 
& Medium (O(n/log n) compression) 
& Compression may degrade tail accuracy \\
\hline

RAOP ~\cite{RAOP}  & RNN-LSTM  
& Not specified 
& 3.22$\times$ accuracy; 4.2$\times$ coverage 
& 15.6\% efficiency gain 
& Low (significant compute overhead) 
& Requires per-workload training; prediction delay \\
\hline

Voyager ~\cite{voyager}  &  Hierarchical NN  
& Moderate (multi-level) 
& 19.4\% higher coverage vs.\ ISB 
& Significant 
& Medium 
& Complex multi-level architecture; training data-intensive \\
\hline

Pathfinder ~\cite{b10} &  Spiking NN (SNN)  
& Low (event-driven) 
& Competitive with conventional NN 
& Competitive 
& High (bio-inspired; energy-efficient) 
& Emerging paradigm; limited benchmarking \\
\hline

PPF~\cite{PPF}  &  Perceptron filter  
& $\sim$13 KB (perceptron tables) 
& +3.78\% (1 core); +11.4\% (4 core) over SPP 
& 3.78–11.4\% over SPP 
& High (simple linear model; on-chip feasible) 
& Struggles with highly dynamic workloads \\
\hline

Pythia ~\cite{pythia}  & Deep RL (Q-table)  
& 1.03\% chip area 
& +3.4–3.8\% (1 core); 16.9–20.2\% (BW-constrained) 
& 3.4–20.2\% over MLOP/Bingo 
& High (config registers; no SW changes) 
& Learning warm-up phase; limited system feedback \\
\hline

Bhattacharjee et al. ~\cite{ref25} &  GNN  
& High (graph construction) 
& Effective for graph workloads 
& Reported gains on graph analytics 
& Low (training overhead) 
& Graph-specific; poor generalization \\
\hline

Alcorta et al.~\cite{ref11} &  Decision Tree   
& 42 bytes 
& Up to 25\% speedup 
& Up to 25\% 
& Very high (trivial on-chip cost) 
& Sensitive to training data; no per-core tuning \\
\hline

Wang et al. ~\cite{ref15} &  CNN-LSTM  
& Not specified 
& Superior to LRU/LFU baselines 
& Improved response time 
& Low (dual-network pipeline) 
& Simulated only; pipeline latency concerns \\
\hline

SeLeP ~\cite{SeLaP}  &  LSTM + dynamic partitioning   
& Moderate 
& Up to 40\% hit rate increase; avg 95\% hit rate 
& Up to 45\% I/O wait reduction 
& Medium (DB-level integration) 
& Tuned for exploratory DB workloads; unclear generalization \\
\hline

\end{tabular}

\caption{Comprehensive Comparison of ML-Based Prefetching Techniques}
\label{tab:ml_prefetch_detailed}
\end{table*}
\subsubsection{Synthesis: Training Cost, Inference Latency, Hardware Feasibility, and Generalization}
The comparative analysis reveals that the practical viability of ML-based prefetchers hinges on four interrelated dimensions: training cost, inference latency, hardware feasibility, and generalization ability. This section synthesizes findings across all reviewed techniques along these axes (summarized in Table 7) and identifies the critical design tensions that shape future research directions.
     \subsubsection*{Training Cost:}
The divide between online and offline training fundamentally shapes the deployment model. Offline approaches such as Delta-LSTM ~\cite{Delta-LSTM}  and Voyager ~\cite{voyager}  require collecting representative memory traces, training on GPU infrastructure, and embedding frozen weights into the prefetcher. This pipeline introduces significant upfront cost and creates a dependency on trace representativeness: if the production workload diverges from training traces, accuracy degrades. RAOP ~\cite{RAOP}  explicitly notes the need for per-workload retraining. In contrast, online approaches like PPF~\cite{PPF}  and Pythia ~\cite{pythia}  update weights or Q-values incrementally during execution. PPF's perceptron updates are essentially single-cycle operations, while Pythia's RL updates occur per demand request. The cost is a learning warm-up phase—during the initial execution period, the model operates with random or suboptimal parameters. Pathfinder ~\cite{b10} occupies an interesting middle ground: its STDP-based learning is online and event-driven, but the convergence behavior of spiking networks in the prefetching context is not yet well characterized. The practical implication is that online models are better suited for production deployment where workloads shift over time, while offline models are more appropriate for simulation-guided design exploration or fixed-workload scenarios.
 \subsubsection*{Inference Latency:}
Inference latency is arguably the most critical constraint for on-chip prefetchers, since the prediction must be issued before the demand access arrives. Perceptron models (PPF) compute a simple dot product requiring 1–3 cycles. Pythia's RL lookup involves indexing a Q-table, feasible in approximately 5 cycles. Both are compatible with L1/L2 cache timings. Decision tree traversal (Alcorta et al. ~\cite{ref11}) is comparable to a branch prediction lookup—effectively a single cycle. In stark contrast, a full LSTM forward pass involves matrix multiplications across multiple gates, requiring tens to hundreds of cycles even with aggressive hardware optimization. This makes uncompressed LSTM models unsuitable for latency-critical cache levels. Compressed-LSTM ~\cite{CompressedLSTM} alleviates this partially through its O(n/log n) output encoding, but the recurrent computation itself remains expensive. GNN inference (Bhattacharjee et al. ~\cite{ref25}) additionally requires graph construction and message-passing steps that add further latency. The CNN-LSTM hybrid of Wang et al. ~\cite{ref15} compounds the issue by chaining two inference pipelines. This analysis suggests a natural architectural mapping: lightweight online models for L1/L2 prefetching where latency budgets are tight (< 10 cycles), and more expressive offline models for LLC or memory-side prefetching where latency tolerance is higher and the cost of a cache miss is orders of magnitude greater.
 \subsubsection*{Hardware Feasibility:}
On-chip storage budgets for prefetcher metadata are typically constrained to 8–16 KB in commercial processors. This constraint immediately disqualifies uncompressed LSTM models (~4 MB for Delta-LSTM) from on-chip deployment. Compressed-LSTM reduces this to ~40 KB—still above typical budgets but within reach for dedicated accelerators or memory-side logic. The most hardware-amenable designs are PPF (~13 KB), Pythia (1.03\% chip area, dominated by the Q-table), and the decision tree of Alcorta et al. (42 bytes). Pathfinder's spiking neural network is event-driven and inherently sparse, but its actual silicon implementation cost remains to be validated. It is worth noting that hardware feasibility extends beyond storage: power consumption, thermal constraints, and integration with existing cache controller logic all affect practical deployment. Pythia's design is notable for its use of simple configuration registers that allow the RL objective to be customized without hardware changes—a form of post-silicon tunability that is valuable in commercial settings. The GNN and CNN-LSTM approaches have not demonstrated hardware implementations and currently operate only in simulation, limiting assessment of their true feasibility.
 \subsubsection*{Generalization Ability:}Generalization—the ability to maintain prediction quality across diverse and unseen workloads—remains the most open challenge. Offline supervised models are inherently workload-specific: Delta-LSTM, Voyager, and RAOP train on fixed trace distributions and degrade when access patterns diverge from training data. Bhattacharjee et al.'s GNN is effective for graph analytics but offers limited utility for non-graph workloads. SeLeP ~\cite{SeLaP}  is tailored to exploratory database queries and its generalization to general-purpose CPU workloads is unclear. Online models fare better in principle—PPF and Pythia adapt to the current workload at runtime—but their limited model capacity (linear perceptron, shallow Q-table) means they cannot capture the deep, non-linear correlations that LSTM models handle. This creates a fundamental tension between adaptability and model capacity that current designs do not resolve.
A promising direction emerging from this analysis is hierarchical or ensemble strategies that combine the strengths of different paradigms. For instance, a lightweight online model (decision tree or perceptron) could act as a first-stage selector or filter, dynamically choosing among or gating multiple specialized models. The 42-byte decision tree of Alcorta et al. already demonstrates this concept at the prefetcher-selection level. Extending this to a finer-grained model-switching or model-ensembling framework—where, say, a perceptron handles regular stride phases while an LSTM variant activates during irregular phases—could achieve both broad generalization and deep pattern capture without imposing the overhead of the most expensive model at all times. However, such architectures introduce coordination complexity and remain largely unexplored in the prefetching literature.


\section{Research Gaps and Future Directions}
\label{sec:research-gaps}

The preceding sections have surveyed the landscape of hardware, software, hybrid, and ML-based prefetching techniques for complex memory access patterns. While substantial progress has been made---from early stride detectors to reinforcement learning agents---our analysis reveals several persistent gaps that limit the practical impact and broader applicability of current approaches. This section systematically enumerates these gaps, organized into seven thematic categories, and identifies concrete future research directions for each.

\subsection{Standardized Evaluation and Benchmarking}
\label{sec:gap-benchmarks}

\textbf{Gap 1: Absence of standardized benchmarks for ML-based prefetchers.}
A persistent obstacle in evaluating ML-based prefetching is the lack of a unified benchmarking framework. Existing studies employ heterogeneous evaluation setups: Delta-LSTM~\cite{Delta-LSTM} and Voyager~\cite{voyager} use subsets of SPEC CPU traces with different simulation configurations, Pythia~\cite{pythia} evaluates on ChampSim with a specific multi-core setup, and SeLeP~\cite{SeLaP} targets database workloads on entirely different platforms. This fragmentation makes direct comparison across techniques unreliable. Unlike the branch prediction community, which benefits from the Championship Branch Prediction (CBP) series with standardized traces, metrics, and simulation infrastructure, the prefetching community lacks an equivalent unified competition framework that covers both regular and irregular workloads. The Data Prefetching Championship (DPC) series has made partial progress but primarily targets conventional prefetchers and does not yet incorporate ML-specific evaluation criteria such as training cost, inference latency, or generalization across trace distributions.

\textit{Future direction:} Development of a comprehensive prefetching benchmark suite that includes (i)~diverse workload categories (server, HPC, graph analytics, database, mobile), (ii)~standardized simulation infrastructure with fixed cache hierarchy configurations, (iii)~ML-specific metrics (training time, inference cycles, model size, energy per prediction), and (iv)~cross-workload generalization tests where models trained on one workload class are evaluated on another. Such a benchmark would accelerate progress by enabling reproducible, apples-to-apples comparisons.

\textbf{Gap 2: Inconsistent metric reporting across studies.}
As discussed in Section~2.1, prefetcher evaluation relies on metrics such as coverage, accuracy, timeliness, IPC improvement, and MPKI reduction. However, our survey reveals that different studies report different subsets of these metrics, often under different definitions. For instance, some works report coverage as the fraction of L2 misses eliminated, while others report it relative to LLC misses. IPC improvements are reported against different baselines (no prefetcher, stride prefetcher, or a specific prior technique). This inconsistency hinders meta-analysis and makes it difficult to assess whether the field is making genuine progress or merely optimizing for different measurement criteria.

\textit{Future direction:} The community should converge on a minimal reporting standard---a core set of metrics (coverage, accuracy, IPC speedup, MPKI reduction, memory traffic overhead, and storage cost) reported against a common no-prefetcher baseline and at least one established prefetcher baseline (e.g., BOP or SPP). Journals and conferences could encourage or mandate such reporting through reviewer guidelines.

\subsection{Prefetching for Emerging Memory Architectures}
\label{sec:gap-architectures}

\textbf{Gap 3: Prefetching for heterogeneous and near-memory computing.}
Near-memory and processing-in-memory (PIM) architectures place compute units adjacent to or within memory arrays, fundamentally altering the data movement economics. While ContextPrefetcher~\cite{b15} addresses near-storage accelerators, prefetching for PIM architectures (e.g., UPMEM, Samsung HBM-PIM, etc.) remains largely unexplored. The challenge is that traditional prefetchers assume a clear processor--memory hierarchy, whereas PIM blurs this boundary: the ``processor'' is inside the memory, and the relevant prefetching target may be intra-bank data movement rather than cache-line fetches.

\textit{Future direction:} Design of prefetching mechanisms tailored to PIM architectures, potentially involving intra-bank or intra-vault data staging guided by access pattern prediction. The lightweight, low-storage models (perceptron, decision tree) identified in Section~6.2 are particularly relevant here, given the severe area and power constraints of in-memory compute units.

\subsection{Energy Efficiency and Sustainability}
\label{sec:gap-energy}

\textbf{Gap 4: Inadequate characterization of energy--accuracy trade-offs in neural prefetchers.}
Energy consumption is a first-order design constraint in modern processors, yet the energy overhead of ML-based prefetchers is poorly characterized. Of the eleven ML-based techniques reviewed in Section~6.2, only Pythia~\cite{pythia} reports chip area overhead (1.03\%), and none provide detailed energy-per-prediction measurements. The energy cost of an LSTM forward pass, a perceptron dot product, or an RL Q-table lookup at the hardware level has not been rigorously quantified in the prefetching literature. This gap is particularly concerning because the energy saved by reducing cache misses must be weighed against the energy consumed by the prediction mechanism itself. A prefetcher that improves IPC by 5\% but increases total chip power by 10\% may be counterproductive in power-constrained environments such as mobile SoCs or data center processors operating under thermal limits.

\textit{Future direction:} Rigorous energy modeling of ML-based prefetchers using industry-standard tools (e.g., McPAT, CACTI, or post-synthesis power analysis). Studies should report energy per prediction, total prefetcher power as a fraction of core power, and the net energy impact (energy saved from reduced misses minus energy consumed by prediction). Spiking neural networks (Pathfinder~\cite{b10}) and event-driven architectures warrant particular attention as potentially energy-efficient alternatives to conventional neural models.

\textbf{Gap 5: Sustainability implications of training overhead for offline ML prefetchers.}
Offline ML-based prefetchers require GPU-hours of training on memory traces, and this training must be repeated whenever the target workload changes. As processors support increasingly diverse workloads, the cumulative training energy cost becomes non-trivial. This concern parallels the broader sustainability discussion in the ML community regarding the carbon footprint of model training, but has not been addressed in the prefetching literature.

\textit{Future direction:} Investigation of transfer learning, few-shot adaptation, and meta-learning techniques that allow offline prefetching models trained on one workload class to be rapidly fine-tuned for new workloads with minimal additional training. This would reduce the amortized training energy cost and make offline models more practical for heterogeneous deployment scenarios.

\subsection{Prefetching for Emerging Workloads}
\label{sec:gap-workloads}

\textbf{Gap 6: Prefetching for large language model (LLM) inference workloads.}
The rapid deployment of transformer-based large language models has introduced a new class of memory-intensive workloads with distinctive access patterns: massive embedding table lookups with irregular indices, key-value (KV) cache accesses that grow dynamically with sequence length, and attention computations that produce access patterns dependent on input content. These patterns are qualitatively different from the SPEC CPU, graph analytics, and database workloads on which existing prefetchers are evaluated. The memory bottleneck in LLM inference---particularly the memory-bound decode phase---makes it a prime target for intelligent prefetching, yet no prefetching study in our survey specifically addresses LLM inference access patterns.

\textit{Future direction:} Characterization of memory access patterns in LLM inference (prefill vs.\ decode phases, KV cache access patterns across attention heads, embedding table access distributions) followed by the development of workload-specific prefetching strategies. The semantic prefetching paradigm (Section~4.2) may be particularly relevant, as the access patterns in LLM inference carry rich semantic structure (e.g., attention patterns reflect linguistic relationships).

\textbf{Gap 7: Prefetching for GNN and graph analytics at scale.}
While Bhattacharjee et al.~\cite{ref25} have applied GNNs to learn memory access patterns in graph workloads, and AMC~\cite{ref27} targets evolving graph analytics, prefetching for large-scale GNN training and inference remains underexplored. GNN workloads combine the irregular access patterns of graph traversals with the regular patterns of neural network computation, creating a hybrid access pattern that neither purely spatial nor purely temporal prefetchers handle well. Additionally, GNN workloads operating on billion-edge graphs exhibit access patterns that span multiple memory tiers and may exceed the capacity of on-chip metadata structures.

\textit{Future direction:} Development of hybrid prefetchers that dynamically switch between spatial strategies (for the neural network computation phases) and temporal/correlation strategies (for the graph sampling and message-passing phases). The phase-based prefetcher selection approach of Alcorta et al.~\cite{ref11} could serve as a starting point for such dynamic switching.

\textbf{Gap 8: Prefetching for sparse and irregular scientific computing.}
Sparse matrix operations (SpMV, SpGEMM) and irregular stencil computations are central to scientific computing and remain challenging for prefetchers due to index-dependent access patterns. While VLDP~\cite{VLDP} and ISB~\cite{Temporal4} address some forms of irregularity, the interaction between sparse storage formats (CSR, CSC, COO, block-sparse) and prefetcher effectiveness has not been systematically studied. Different sparse formats produce fundamentally different access patterns for the same underlying computation, and the optimal prefetching strategy may depend on the storage format rather than the algorithm alone.

\textit{Future direction:} Format-aware prefetching that leverages knowledge of the sparse storage format (potentially communicated via software hints or ISA extensions) to predict access patterns. This represents a natural application of the hybrid software-guided hardware prefetching paradigm discussed in Section~5.2.

\subsection{Security Implications of Prefetching}
\label{sec:gap-security}

\textbf{Gap 9: Prefetching as a side-channel attack vector.}
Hardware prefetchers create observable microarchitectural side effects---changes in cache state, memory bus activity, and timing---that can be exploited by adversaries to infer victim access patterns. Spectre-class attacks have demonstrated that speculative execution can be weaponized, and prefetchers share a similar risk profile: they bring data into the cache based on predicted access patterns, potentially making those predictions observable through cache timing side channels. Despite this well-known risk, none of the prefetching techniques reviewed in this survey discuss security implications, and the ML-based prefetchers introduce additional concerns. If an attacker can influence the training data or reward signals of an online ML prefetcher (e.g., Pythia or PPF), they may be able to manipulate the prefetcher's behavior to leak information about co-located processes.

\textit{Future direction:} Security analysis of modern prefetchers, particularly ML-based designs, under adversarial threat models. Research should investigate (i)~whether ML prefetcher training can be poisoned to create covert channels, (ii)~whether the prefetcher's learned model itself leaks information about the workload it was trained on (model inversion attacks), and (iii)~design of security-aware prefetchers that bound information leakage while maintaining performance benefits. Partitioned or isolated prefetcher state per security domain is one potential mitigation.

\subsection{Hardware--Software Co-Design and Compiler Integration}
\label{sec:gap-codesign}

\textbf{Gap 10: Limited compiler support for ML-aware prefetching.}
The hybrid approaches reviewed in Section~5 demonstrate the value of software guidance for hardware prefetchers. However, current compiler support is largely limited to inserting traditional \texttt{\_\_builtin\_prefetch} intrinsics or annotating code regions (as in RnR~\cite{RnR} and GRP~\cite{GRP}). No existing compiler framework generates annotations specifically tailored to ML-based hardware prefetchers---for example, providing phase boundary hints to an RL prefetcher, or communicating data structure type information to a GNN-based predictor. DTAP~\cite{DTAP} takes a step in this direction by using compiler-extracted type information, but its integration is limited to strongly-typed languages and specific data structures.

\textit{Future direction:} Development of compiler passes that generate rich metadata for ML-based hardware prefetchers, including phase boundaries, data structure types, expected access pattern classes (stride, pointer-chase, random), and workload-specific hints. This metadata could be communicated via ISA extensions, memory-mapped configuration registers, or a dedicated hint interface. The LLVM infrastructure, with its extensive analysis passes, is a natural platform for such development.

\textbf{Gap 11: Cross-layer prefetching across the full stack.}
Current prefetching techniques typically operate at a single layer: hardware prefetchers at the cache controller, compiler-based prefetchers at the instruction level, OS-level prefetchers at the page level (e.g., HoPP~\cite{hopp}), or application-level prefetchers within specific frameworks (e.g., SeLeP~\cite{SeLaP} for databases). CrossPrefetch~\cite{CrossPrefetch} demonstrates the benefits of OS-user cooperation for I/O prefetching, but a truly cross-layer prefetching framework that coordinates decisions across hardware, OS, compiler, and application layers does not exist. Each layer has access to different information: the application knows data semantics, the compiler knows control flow, the OS knows page-level locality, and the hardware knows cache-line-level timing. No single-layer prefetcher can access all of this information.

\textit{Future direction:} Design of a cross-layer prefetching interface that enables information flow between stack layers. For example, the application could communicate expected access patterns to the OS, which relays page-level hints to the hardware prefetcher while also adjusting its own page prefetching. Such a framework would require standardized interfaces (potentially extending CXL's coherence protocols or leveraging RISC-V's custom instruction space) and a coordination mechanism to avoid conflicting prefetch decisions across layers.

\textbf{Gap 12: Interpretability of ML-based prefetching models.}
ML-based prefetchers, particularly deep neural models (LSTM, GNN, CNN-LSTM), operate as black boxes: they produce predictions but do not explain why a particular address was predicted. This lack of interpretability creates practical challenges. When a neural prefetcher degrades performance on a particular workload phase, designers cannot diagnose whether the degradation is due to a distribution shift, insufficient model capacity, or a fundamental property of the access pattern. Perceptron-based models (PPF) offer partial interpretability through their learned feature weights, but deeper models remain opaque.

\textit{Future direction:} Application of ML interpretability techniques (attention visualization, feature attribution, SHAP values) to prefetching models to understand what program features and access patterns they have learned to exploit. Such analysis could also reveal whether different ML architectures converge on similar internal representations of access patterns, potentially informing the design of more efficient, purpose-built prediction circuits.

\subsection{Summary}
\label{sec:gaps-summary}

The twelve research gaps identified above span the full spectrum from foundational theory (Gap~12) to immediate engineering needs (Gap~1). Two themes emerge as particularly urgent. First, the \textbf{evaluation infrastructure} for ML-based prefetchers is inadequate: without standardized benchmarks, consistent metrics, and energy characterization, the community cannot reliably measure progress. Second, \textbf{emerging workloads}---particularly LLM inference and large-scale GNN computation---present access patterns that are qualitatively different from the benchmarks on which existing prefetchers were designed and evaluated. Addressing these gaps will require coordinated effort across the architecture, compiler, and machine learning communities, and we hope that this survey provides a useful roadmap for that effort.

\section{Conclusion}
\label{sec:conclusion}

This survey has presented a comprehensive review of memory access prediction techniques for CPU cache prefetching, spanning hardware-based, software-based, hybrid, and machine learning-driven approaches. By systematically examining over 115 works published between 1991 and 2025, we have traced the evolution of prefetching from simple next-line and stride detectors to sophisticated reinforcement learning agents and deep neural predictors, while identifying the persistent challenges and emerging opportunities that define the current research landscape.

\paragraph{Key contributions of this survey.}
First, we proposed a structured taxonomy that organizes prefetching techniques along multiple dimensions: spatial versus temporal locality exploitation (Section~3), hardware versus software versus hybrid implementation (Sections~3--5), and critically for the ML-based techniques that motivate the survey's title---supervised versus reinforcement versus unsupervised learning paradigms with online versus offline training modes (Section~6.2.1). The comprehensive comparison of ML-Based techniques summarized in Table~\ref{tab:ml_prefetch_detailed}, provides researchers with a principled framework for positioning new techniques relative to existing work.

Second, we provided a multi-dimensional comparative analysis of eleven ML-based prefetchers, evaluated across storage overhead, accuracy, IPC improvement, hardware feasibility, and key limitations. This analysis revealed three cross-cutting findings: (i)~a clear accuracy--overhead Pareto frontier defined by model class, where perceptron and RL models occupy the low-overhead region and LSTM-based models occupy the high-accuracy region; (ii)~a natural architectural mapping between model complexity and cache hierarchy level, where lightweight online models suit latency-critical L1/L2 caches and more expressive offline models suit LLC or memory-side deployment; and (iii)~a fundamental tension between adaptability and model capacity that no single current technique resolves, motivating hierarchical or ensemble architectures as a key open direction.

Third, we identified twelve concrete research gaps organized across five thematic categories (Section~\ref{sec:research-gaps}), ranging from the absence of standardized ML-prefetcher benchmarks (Gap~1) and the lack of prefetching studies targeting LLM inference workloads (Gap~6) to the need for information-theoretic frameworks for analyzing prefetchability (Gap~12). These gaps provide a structured roadmap for future research.

\paragraph{Broader observations.}
Several broader trends emerge from this survey that are worth highlighting for researchers and system designers.

The field is converging toward \textit{intelligence at the edge of the memory hierarchy}. Early prefetchers relied on fixed heuristics; modern designs increasingly embed learning capability---whether through perceptron filters, reinforcement learning agents, or neural predictors---directly into cache controller logic. This trend is likely to accelerate as workloads become more irregular and data-dependent, making static prediction rules increasingly inadequate.

However, \textit{the gap between research prototypes and deployed systems remains wide}. Of the ML-based prefetchers we reviewed, only PPF, Pythia, and the decision tree selector of Alcorta et al.~\cite{ref11} have demonstrated feasibility at realistic on-chip resource budgets. The majority of neural prefetchers (LSTM, GNN, CNN-LSTM variants) operate exclusively in simulation, and their true hardware costs, including power, area, thermal impact, and integration complexity, remain uncharacterized. Bridging this simulation-to-silicon gap is perhaps the single most important challenge for the field.

The \textit{architectural landscape is shifting} in ways that invalidate the assumptions underlying most existing prefetchers. Processing-in-memory architectures introduce variable latencies, shared bandwidth pools, and blurred processor memory boundaries that conventional prefetchers were not designed to handle. Simultaneously, emerging workloads, particularly large language model inference and large-scale graph neural network computation, present access patterns that are qualitatively different from the SPEC CPU and server benchmarks on which the vast majority of prefetchers have been evaluated.

Finally, \textit{cross-layer coordination represents an underexplored frontier}. Current prefetching techniques operate overwhelmingly in a single layer of the system stack. Yet the information needed for optimal prefetching is distributed across the application (which knows data semantics), the compiler (which knows control flow), the operating system (which knows page-level locality), and the hardware (which knows cache-line-level timing). No existing framework coordinates prefetching decisions across these layers, and designing such a framework with standardized interfaces and conflict resolution mechanisms---is a substantial systems research challenge.

\paragraph{Closing remarks.}
Data prefetching has been a cornerstone of processor performance optimization for over four decades. As we enter an era of increasingly heterogeneous architectures, irregular workloads, and intelligent hardware, the prefetching problem is not becoming simpler it is becoming richer. The transition from pattern matching to pattern learning, from fixed heuristics to adaptive agents, and from single-layer optimization to cross-stack coordination defines the next chapter of prefetching research. We hope this survey provides both a useful reference for the current state of the art and a clear roadmap for the challenges and opportunities that lie ahead.
\bibliographystyle{unsrt}
\bibliography{sample-base}
\end{document}